\documentclass[12pt]{article}
\PassOptionsToPackage{hyphens}{url}
\usepackage[pagebackref=true,colorlinks]{hyperref}
\usepackage{amsmath,amssymb,amsthm,mathtools,amsrefs,mathrsfs}
\usepackage{pifont}
\usepackage{mathpazo} 
\usepackage{eulervm}
\usepackage{csquotes}
\usepackage{bbm}
\usepackage{lscape}
\usepackage[table]{xcolor}
\usepackage{tikz}
\usetikzlibrary{arrows.meta}
\usetikzlibrary{decorations.pathmorphing,shapes}
\usetikzlibrary{calc}
\usepackage{comment}
\usepackage{adjustbox}
\usepackage{scalerel,stackengine}
\usepackage{stmaryrd}
\usepackage{fancyhdr}
\usepackage{soul}
\usepackage{dsfont}
\usepackage[normalem]{ulem}
\usepackage{longtable}
\usepackage[bottom]{footmisc}
\usepackage{graphicx}
\usepackage{caption}
\usepackage{subcaption}
\usepackage{pictexwd, dcpic}
\usepackage{float}
\usepackage{enumerate}
\usepackage[all]{xy} 
\usetikzlibrary{circuits.ee.IEC}
\hypersetup{
	colorlinks=true,
    linktoc=all,
    linkcolor=blue,  
    citecolor=blue,
    }

\usepackage{tocbasic}
\DeclareTOCStyleEntry[
  beforeskip=-0.3em plus 1pt,
  pagenumberformat=\textbf
]{tocline}{section}

\makeatletter
\newcommand*{\shifttext}[2]{%
  \settowidth{\@tempdima}{#2}%
  \makebox[\@tempdima]{\hspace*{#1}#2}%
}
\makeatother
\makeatletter
\renewcommand*\env@matrix[1][\arraystretch]{%
  \edef\arraystretch{#1}%
  \hskip -\arraycolsep
  \let\@ifnextchar\new@ifnextchar
  \array{*\c@MaxMatrixCols c}}
\makeatother

\stackMath
\newcommand\reallywidehat[1]{%
\savestack{\tmpbox}{\stretchto{%
  \scaleto{%
    \scalerel*[\widthof{\ensuremath{#1}}]{\kern.1pt\mathchar"0362\kern.1pt}%
    {\rule{0ex}{\textheight}}
  }{\textheight}%
}{2.4ex}}%
\stackon[-6.9pt]{#1}{\tmpbox}%
}
\makeatletter
\pgfmathdeclarefunction{erf}{1}{%
  \begingroup
    \pgfmathparse{#1 > 0 ? 1 : -1}%
    \edef\sign{\pgfmathresult}%
    \pgfmathparse{abs(#1)}%
    \edef\x{\pgfmathresult}%
    \pgfmathparse{1/(1+0.3275911*\x)}%
    \edef\t{\pgfmathresult}%
    \pgfmathparse{%
      1 - (((((1.061405429*\t -1.453152027)*\t) + 1.421413741)*\t 
      -0.284496736)*\t + 0.254829592)*\t*exp(-(\x*\x))}%
    \edef\y{\pgfmathresult}%
    \pgfmathparse{(\sign)*\y}%
    \pgfmath@smuggleone\pgfmathresult%
  \endgroup
}
\makeatother

\theoremstyle{plain}
\newtheorem{theorem}[equation]{Theorem}
\newtheorem{lemma}[equation]{Lemma}
\newtheorem{proposition}[equation]{Proposition}
\newtheorem{corollary}[equation]{Corollary}
\theoremstyle{definition}
\newtheorem{definition}[equation]{Definition}
\newtheorem{construction}[equation]{Construction}
\newtheorem{question}[equation]{Question}

\newtheorem{problem}[equation]{Problem}
\newtheorem{example}[equation]{Example}
\newtheorem{exercise}[equation]{Exercise}
\newtheorem*{answer}{Answer}
\newtheorem*{solution}{Solution}
\newtheorem{remark}[equation]{Remark}

\newtheorem{notation}[equation]{Notation}
\newtheorem{noterm}[equation]{Notation and Terminology}

\newcommand\define[1]{\emph{\textbf{#1}}}

\numberwithin{equation}{section}

   \let\de=\delta

 \let\t=\tau

\let\C=\Chi

\newcommand{\be}{\begin{equation}}
\newcommand{\ee}{\end{equation}}
\def\ba{\begin{align}} 
\def\ea{\end{align}}
\newcommand{\bea}{\begin{eqnarray}}
\newcommand{\eea}{\end{eqnarray}}
\newcommand{\bx}{\begin{example}}
\newcommand{\ex}{\end{example}}
\newcommand{\bex}{\begin{exercise}}
\newcommand{\eex}{\end{exercise}}
\newcommand{\ban}{\begin{answer}}
\newcommand{\ean}{\end{answer}}
\newcommand{\bt}{\begin{theorem}}
\newcommand{\et}{\end{theorem}}
\newcommand{\bc}{\begin{corollary}}
\newcommand{\ec}{\end{corollary}}
\newcommand{\blem}{\begin{lemma}}
\newcommand{\elem}{\end{lemma}}
\newcommand{\bp}{\begin{problem}}
\newcommand{\ep}{\end{problem}}
\newcommand{\bn}{\begin{proposition}}
\newcommand{\en}{\end{proposition}}
\newcommand{\bd}{\begin{definition}}
\newcommand{\ed}{\end{definition}}
\newcommand{\bcon}{\begin{construction}}
\newcommand{\econ}{\end{construction}}
\newcommand{\bq}{\begin{question}}
\newcommand{\eq}{\end{question}}
\newcommand{\bprf}{\begin{proof}}
\newcommand{\eprf}{\end{proof}}
\newcommand{\br}{\begin{remark}}
\newcommand{\er}{\end{remark}}
\newcommand{\bs}{\begin{solution}}
\newcommand{\es}{\end{solution}}
\newcommand{\beqs}{\begin{eqnarray}}
\newcommand{\eeqs}{\end{eqnarray}}
\newcommand{\bnt}{\begin{noterm}}
\newcommand{\ent}{\end{noterm}}
\newcommand{\bnot}{\begin{notation}}
\newcommand{\enot}{\end{notation}}

\newcommand{\<}{\langle}
\renewcommand{\>}{\rangle}

\newcommand{\id}{\mathrm{id}}

\newcommand{\mC}{\mathcal{C}}

\newcommand{\tr}{{\rm tr} }

\def\R{{{\mathbb R}}}
\def\C{{{\mathbb C}}}

\def\mA{{{\mathcal{A}}}}

\def\mB{{{\mathcal{B}}}}

\def\mZ{{{\mathcal{Z}}}}
\newcommand{\matr}{\mathbb{M}}

\newcommand{\sahom}{\mathbf{Hom}^{\mathrm{sa}}}
\renewcommand{\hom}{\mathbf{Hom}}
\newcommand{\CPTP}{\mathbf{CPTP}}

\def\invexcl{\rotatebox[origin=c]{180}{$!$}}
\newcommand{\bloom}{\operatorname{\invexcl}}
\newcommand{\Lbloom}[1]{\sideset{_{#1}}{}\bloom}

\newcommand{\shriek}{\operatorname{!}}
\newcommand{\Cs}{\operatorname{\mathscr{S}}}
\newcommand{\Cd}{\operatorname{\mathscr{D}}}

\def\VA{\mathcal{A}}
\def\VB{\mathcal{B}}
\def\VC{\mathcal{C}}

\def\M{\mathbb{M}}

\makeatletter
\newcommand{\ostar}{\mathbin{\mathpalette\make@circled\star}}
\newcommand{\make@circled}[2]{%
  \ooalign{$\m@th#1\smallbigcirc{#1}$\cr\hidewidth$\m@th#1#2$\hidewidth\cr}%
}
\newcommand{\smallbigcirc}[1]{%
  \vcenter{\hbox{\scalebox{0.77778}{$\m@th#1\bigcirc$}}}%
}
\makeatother

\newcommand{\xrightarrowdbl}[2][]{%
  \xrightarrow[#1]{#2}\mathrel{\mkern-14mu}\rightarrow
}

\newcommand{\stoch}{\;\xy0;/r.25pc/:(-3,0)*{}="1";(3,0)*{}="2";{\ar@{~>}"1";"2"|(1.06){\hole}};\endxy\!}
\newcounter{sarrow}

\newcounter{sqarrow}

\newcommand{\ben}{\renewcommand{\theenumi}{\alph{enumi}} 
\renewcommand{\labelenumi}{(\theenumi)}\begin{enumerate}}
\newcommand{\een}{\end{enumerate}}

\newcommand\blfootnote[1]{%
  \begingroup
  \renewcommand\thefootnote{}\footnote{#1}%
  \addtocounter{footnote}{-1}%
  \endgroup
}

\newlength\stateheight
\newlength\minimumstatewidth
\tikzset{width/.initial=\minimummorphismwidth}
\tikzset{colour/.initial=white}

\newif\ifblack\pgfkeys{/tikz/black/.is if=black}
\newif\ifwedge\pgfkeys{/tikz/wedge/.is if=wedge}
\newif\ifvflip\pgfkeys{/tikz/vflip/.is if=vflip}
\newif\ifhflip\pgfkeys{/tikz/hflip/.is if=hflip}
\newif\ifhvflip\pgfkeys{/tikz/hvflip/.is if=hvflip}

\pgfdeclarelayer{foreground}
\pgfdeclarelayer{background}
\pgfsetlayers{background,main,foreground}

\def\thickness{0.4pt}

\makeatletter
\pgfkeys{%
  /tikz/on layer/.code={
    \pgfonlayer{#1}\begingroup
    \aftergroup\endpgfonlayer
    \aftergroup\endgroup
  },
  /tikz/node on layer/.code={
    \gdef\node@@on@layer{%
      \setbox\tikz@tempbox=\hbox\bgroup\pgfonlayer{#1}\unhbox\tikz@tempbox\endpgfonlayer\pgfsetlinewidth{\thickness}\egroup}
    \aftergroup\node@on@layer
  },
  /tikz/end node on layer/.code={
    \endpgfonlayer\endgroup\endgroup
  }
}
\def\node@on@layer{\aftergroup\node@@on@layer}
\makeatother

\makeatletter
\pgfdeclareshape{state}
{
    \savedanchor\centerpoint
    {
        \pgf@x=0pt
        \pgf@y=0pt
    }
    \anchor{center}{\centerpoint}
    \anchorborder{\centerpoint}
    \saveddimen\overallwidth
    {
        \pgf@x=3\wd\pgfnodeparttextbox
        \ifdim\pgf@x<\minimumstatewidth
            \pgf@x=\minimumstatewidth
        \fi
    }
    \savedanchor{\upperrightcorner}
    {
        \pgf@x=.5\wd\pgfnodeparttextbox
        \pgf@y=.5\ht\pgfnodeparttextbox
        \advance\pgf@y by -.5\dp\pgfnodeparttextbox
    }
    \anchor{A}
    {
        \pgf@x=-\overallwidth
        \divide\pgf@x by 4
        \pgf@y=0pt
    }
    \anchor{B}
    {
        \pgf@x=\overallwidth
        \divide\pgf@x by 4
        \pgf@y=0pt
    }
    \anchor{text}
    {
        \upperrightcorner
        \pgf@x=-\pgf@x
        \ifhflip
            \pgf@y=-\pgf@y
            \advance\pgf@y by 0.4\stateheight
        \else
            \pgf@y=-\pgf@y
            \advance\pgf@y by -0.4\stateheight
        \fi
    }
    \backgroundpath
    {
       \begin{pgfonlayer}{foreground}
        \pgfsetstrokecolor{black}
        \pgfsetlinewidth{\thickness}
        \pgfpathmoveto{\pgfpoint{-0.5*\overallwidth}{0}}
        \pgfpathlineto{\pgfpoint{0.5*\overallwidth}{0}}
        \ifhflip
            \pgfpathlineto{\pgfpoint{0}{\stateheight}}
        \else
            \pgfpathlineto{\pgfpoint{0}{-\stateheight}}
        \fi
        \pgfpathclose
        \ifblack
            \pgfsetfillcolor{black!50}
            \pgfusepath{fill,stroke}
        \else
            \pgfusepath{stroke}
        \fi
       \end{pgfonlayer}
    }
}

\pgfdeclareshape{my ground}
{
  \inheritsavedanchors[from=rectangle ee]
  \inheritanchor[from=rectangle ee]{center}
  \inheritanchor[from=rectangle ee]{north}
  \inheritanchor[from=rectangle ee]{south}
  \inheritanchor[from=rectangle ee]{east}
  \inheritanchor[from=rectangle ee]{west}
  \inheritanchor[from=rectangle ee]{north east}
  \inheritanchor[from=rectangle ee]{north west}
  \inheritanchor[from=rectangle ee]{south east}
  \inheritanchor[from=rectangle ee]{south west}
  \inheritanchor[from=rectangle ee]{input}
  \inheritanchor[from=rectangle ee]{output}
  \anchorborder{\csname pgf@anchor@my ground@input\endcsname}

  \backgroundpath{
    \pgf@process{\pgfpointadd{\southwest}{\pgfpoint{\pgfkeysvalueof{/pgf/outer xsep}}{\pgfkeysvalueof{/pgf/outer ysep}}}}
    \pgf@xa=\pgf@x \pgf@ya=\pgf@y
    \pgf@process{\pgfpointadd{\northeast}{\pgfpointscale{-1}{\pgfpoint{\pgfkeysvalueof{/pgf/outer xsep}}{\pgfkeysvalueof{/pgf/outer ysep}}}}}
    \pgf@xb=\pgf@x \pgf@yb=\pgf@y
    \pgfpathmoveto{\pgfqpoint{\pgf@xa}{\pgf@ya}}
    \pgfpathlineto{\pgfqpoint{\pgf@xa}{\pgf@yb}}
    \pgfmathsetlength\pgf@xa{.5\pgf@xa+.5\pgf@xb}
    \pgfmathsetlength\pgf@yc{.16666\pgf@yb-.16666\pgf@ya}
    \advance\pgf@ya by\pgf@yc
    \advance\pgf@yb by-\pgf@yc
    \pgfpathmoveto{\pgfqpoint{\pgf@xa}{\pgf@ya}}
    \pgfpathlineto{\pgfqpoint{\pgf@xa}{\pgf@yb}}
    \advance\pgf@ya by\pgf@yc
    \advance\pgf@yb by-\pgf@yc
    \pgfpathmoveto{\pgfqpoint{\pgf@xb}{\pgf@ya}}
    \pgfpathlineto{\pgfqpoint{\pgf@xb}{\pgf@yb}}
  }
}
\makeatother

\tikzset{inline text/.style =
  {text height=1.2ex,text depth=0.25ex,yshift=0.5mm}}
\tikzset{arrow box/.style =
  {rectangle,inline text,fill=white,draw,
    minimum height=5mm,yshift=-0.5mm,minimum width=5mm}}
\tikzset{bubble/.style =
  {inner sep=0mm,minimum width=3mm,minimum height=3mm,
    draw,shape=circle,fill=white}}
\tikzset{dot/.style =
  {inner sep=0mm,minimum width=1mm,minimum height=1mm,
    draw,shape=circle}}
\tikzset{white dot/.style = {dot,fill=white,text depth=-0.2mm}}
\tikzset{scalar/.style = {diamond,draw,inner sep=1pt}}
\tikzset{square/.style =
  {inner sep=0mm,minimum width=2mm,minimum height=2mm,
    draw,shape=rectangle}}

\tikzset{star/.style = {dot,fill=white,text depth=-0.2mm}}
\tikzset{copier/.style = {dot,fill,text depth=-0.2mm}}
\tikzset{fakecopier/.style = {square,fill,text depth=-0.2mm}}
\tikzset{discarder/.style = {my ground,draw,inner sep=0pt,
    minimum width=4.2pt,minimum height=11.2pt,anchor=input,rotate=90}}
\tikzset{trace/.style = {my ground,rotate=180,draw,inner sep=0pt, 
    minimum width=4.2pt,minimum height=11.2pt,anchor=input,rotate=90}} 

\tikzset{xshiftu/.style = {shift = {(#1, 0)}}}
\tikzset{yshiftu/.style = {shift = {(0, #1)}}}
\tikzset{scriptstyle/.style={font=\everymath\expandafter{\the\everymath\scriptstyle}}}

\title{On quantum states over time}
\author{James Fullwood and Arthur J.~Parzygnat}

\newcommand{\Addresses}{{
  \bigskip
  \footnotesize

  A.~Parzygnat, \textsc{
  Graduate School of Informatics, Nagoya University, Chikusa-ku, 464-8601 Nagoya, Japan}\par\nopagebreak
  \textit{E-mail address}, A.~Parzygnat: \texttt{parzygnat@nagoya-u.jp}

  \medskip

  J.~Fullwood, \textsc{School of Mathematical Sciences, Shanghai Jiao Tong University, 800 Dongchuan Road, Shanghai, China}\par\nopagebreak
  \textit{E-mail address}, J.~Fullwood: \texttt{fullwood@sjtu.edu.cn}

}}

\begin{document}
\emergencystretch 2em

\maketitle
\begin{abstract}  
In 2017, D.\ Horsman, C.\ Heunen, M.\ Pusey, J.\ Barrett, and R.\ Spekkens proved that there is no physically reasonable assignment that takes a quantum channel and an initial state and produces a joint state on the tensor product of the input and output spaces. 
The interpretation was that there is a clear distinction between space and time in the quantum setting that is not visible classically, where in the latter, one can freely use Bayes' theorem to go between joint states and marginals with noisy channels. 
In this paper, we prove that there actually is such a physically reasonable assignment, bypassing the no-go result of Horsman et al., and we illustrate that this is achievable by restricting the domain of their assignment to a domain which represents the given data more faithfully. 


\end{abstract}

\vspace{-7mm}
\tableofcontents

\section{Introduction}

Given a joint probability measure $p_{XY}$ on the direct product $X\times Y$ of two finite sets, one can obtain the associated marginals $p_{X}$ on $X$ and $p_{Y}$ on $Y$ by pushing these measures forward along the projection maps $\pi_{X}:X\times Y\to X$ and $\pi_{Y}:X\times Y\to Y$, respectively. In addition, one also obtains stochastic maps (i.e., Markov kernels) $p_{Y|X}:X\to Y$ and $p_{X|Y}:Y\to X$, called \emph{conditionals}, such that 
\[
p_{Y|X}(y|x)p_{X}(x)=p_{XY}(x,y)=p_{X|Y}(x|y)p_{Y}(y)
\]
for all $(x,y)\in X\times Y$. This allows one to convert a  joint state, which is a state at a single time, to an initial state together with a stochastic evolution in two distinct ways based on which marginal is used as the initial state. 

Conversely, given a probability measure $p_{X}$ on $X$ and a stochastic map $p_{Y|X}:X\to Y$, one obtains a joint probability measure $p_{XY}$ on $X\times Y$ by the formula 
\[
p_{XY}(x,y):=p_{Y|X}(y|x)p_{X}(x). 
\]
In fact, one can formalize this duality by stating a bijection between these data (modulo some minor subtleties related to measure zero subsets)~\cite{CDDG17}.

As such, one may view a joint state $p_{XY}$ in classical probability either as a state at a single time whose subsystems are arbitrarily separated in space, or as a state \emph{over} time associated with stochastic evolution. Does such a symmetric treatment of space and time hold for quantum systems, namely quantum states and quantum channels? 

The question of when it is possible to go from joint states to marginals and channels was the subject of~\cite{PaQPL21}, though that work only established the conditions needed when the marginals were full rank density matrices and for a particular construction that was motivated by categorical probability theory~\cites{Fr20,ChJa18}. The question of when it is possible to go from initial states and channels to joint states was the subject of the work of Horsman et al.~\cite{HHPBS17}, where they argued that such a construction satisfying a collection of axioms they put forward is not possible. Together, these arguments suggest that the symmetry between time and space that is available (and often taken for granted) in the classical setting might no longer hold for quantum systems and their evolution. 

In this paper, we show that the no-go results of~\cite{HHPBS17} can be bypassed, answering an open question posed at the end of~\cite{HHPBS17}. In particular, we show that \emph{there is} a consistent assignment from quantum channels endowed with initial states to joint states over time that satisfies the axioms proposed in~\cite{HHPBS17}, where the states are represented by self-adjoint, as opposed to positive, density matrices. The way that the no-go result of~\cite{HHPBS17} is bypassed is by restricting the assignment to a domain that reflects the given data more faithfully, rather than demanding a full binary operation as in~\cite{HHPBS17}. Furthermore, we formulate the definitions, axioms, and theorems for arbitrary hybrid classical/quantum systems (i.e., finite-dimensional $C^*$-algebras) and show how these specialize to the setting of purely quantum systems when restricted to matrix algebras. As such, we work in the Heisenberg picture for the formulations of our results, but we translate to the Schr\"odinger picture when specializing to the matrix algebra setting. 

The fact that our state over time does not necessarily correspond to a density matrix, but rather to an observable (self-adjoint matrix), indicates that one cannot in general interpret it as an entity that provides the outcome probabilities of all possible measurements. This is not too surprising since the very act of measuring a \emph{quantum}, as opposed to classical, state influences its statistics at a later point in time as it evolves, such as in the double-slit experiment. Nevertheless, we view this as a feature because when our state over time admits negative eigenvalues, it has been argued in~\cites{FJV15,ZPTGVF18} that the negative eigenvalues can be used to measure temporal correlations. In addition, our construction is closely related to the Jordan product of channels and therefore makes a connection with the resource theory of incompatibility, not only of measurements, but more generally of channels~\cite{GPS21}. Combining these two perspectives with the fact that the degree of compatibility increases when noise is added to the channels leads us to conclude, in a similar fashion to~\cites{FJV15,ZPTGVF18}, that adding noise washes out temporal correlations. In particular, in Remark~\ref{rmk:incompatibility} we show that if sufficient noise is added to a channel and its input state, then the associated state over time given by our proposal becomes a genuine (positive) state. Furthermore, we illustrate in Remark~\ref{rmk:operational} how our proposal of a state over time provides a linear approximation to the state over time proposed by Leifer and Spekkens in~\cites{Le06,Le07,LeSp13}, the latter of which admits an operational interpretation as relating measurements on the output a quantum channel to a probabilistic preparation of states to be sent through the channel. While the state over time proposed by Leifer and Spekkens is locally positive, it lacks desiderata such as compositionality and bilinearity~\cites{LeSp13,HHPBS17}, which our construction achieves.

The main definition of a general states over time function is given in Definition~\ref{defn:statesovertime}. It contrasts with the definition of~\cite{HHPBS17} in that it assumes exactly the data given in the domain rather than assuming that such an assignment extends to a larger domain (for more on this comment, see Remark~\ref{rmk:bypassHHPBS17}). 
The main theorem in this paper is Theorem~\ref{thm:statesovertime}, which provides an explicit construction of a states over time function. At present, it is unknown whether or not there exists other states over time functions differing from our construction.

\section{From quantum channels and quantum states to joint states}

Our results are formulated in the language of finite-dimensional $C^*$-algebras to illustrate the similarities between classical and quantum systems and to include all hybrid classical/quantum systems. Furthermore, we use string diagrams on occasion to provide visualizations of some concepts and proofs, which are sometimes more illuminating than the algebraic manipulations of coordinate expressions. However, such string diagrams are \emph{not} essential to follow the main definitions and statements of results. The string diagrams we use are those of quantum Markov categories~\cite{PaBayes} (in fact, quantum CD/gs-monoidal categories), and the reader is referred to that work for a thorough introduction. A shorter summary of quantum Markov categories is provided in~\cite{PaQPL21}. Classical versions of Markov and CD/gs-monoidal categories originated in the works~\cites{ChJa18,Fr20,Ga96,CoGa99}, which also provide adequate introductions. We assume familiarity with positive and completely positive maps~\cite{Kr83}. 

\begin{notation}
If $m$ is a natural number, then $\matr_{m}(\C)$ denotes the $C^*$-algebra of $m\times m$ matrices with complex entries. The standard matrix units are denoted by $E_{ij}$ (or $E_{ij}^{(m)}$ for additional clarity), while the identity matrix is denoted by $\mathds{1}_{m}$. All $C^*$-algebras in this work will be finite-dimensional and unital, with the involution always written as $\dagger$. As such, all $C^*$-algebras $\mA$ will be \emph{multi-matrix algebras}, i.e., $\mA=\bigoplus_{x\in X}\matr_{m_{x}}(\C)$, where $X$ is a finite set, the $m_{x}$ are natural numbers, and $\dag$ denotes the (component-wise) conjugate transpose. If $\mA$ is a $C^*$-algebra, let $\mu_{\mA}:\mA\otimes\mA\to\mA$ denote the linear product map uniquely determined by sending $A_{1}\otimes A_{2}$ to $A_{1}A_{2}$. The unit in $\mA$ is written as $1_{\mA}$ and the unique unital map from $\C$ to $\mA$ will be denoted by $\shriek_{\mA}$. Meanwhile, $i_{\mA}$ will be used to denote an inclusion of $\mA$ into another algebra with $\mA$ as a tensor factor, such as $\mA\to\mA\otimes\mB$. If $F:\mB\to\mA$ is a linear map, with $\mA=\bigoplus_{x\in X}\matr_{m_{x}}(\C)$ and $\mB=\bigoplus_{y\in Y}\matr_{n_{y}}(\C)$, let $F_{xy}$ denote the $xy$ component of $F$, i.e., the composite $\matr_{n_{y}}(\C)\hookrightarrow\mB\xrightarrow{F}\mA\xrightarrowdbl{\pi_x}\matr_{m_{x}}(\C)$, where $\pi_{x}$ is the projection. Also, let $F^*:\mA\to\mB$ denote the \emph{Hilbert--Schmidt adjoint}, which is the map whose $yx$ component is given by $(F^*)_{yx}=(F_{xy})^*\equiv F_{xy}^*,$ where $F_{xy}^*$ is the usual Hilbert--Schmidt adjoint for linear maps between matrix algebras, namely, it is the unique linear map satisfying 
\[
\tr\big(A_{x}^{\dag}F_{xy}(B_{y})\big)=\tr\big(F^*_{xy}(A_{x})^{\dag}B_{y}\big)
\]
for all $A_{x}\in\matr_{m_{x}}(\C)$ and $B_{y}\in\matr_{n_{y}}(\C)$. 
We will freely use the fact that a linear map is unital if and only if its Hilbert--Schmidt adjoint is trace-preserving. Furthermore, a linear map $F:\mB\to\mA$ is \emph{$\dag$-preserving}, aka \emph{self-adjoint} (meaning $F(B)^{\dag}=F(B^{\dag})$ for all $B\in\mB$), if and only if its Hilbert--Schmidt adjoint is $\dag$-preserving. The vector space of all linear maps from $\mB$ to $\mA$ is denoted by $\hom(\mB,\mA)$, while the affine subspace of $\dag$-preserving maps is denoted by $\sahom(\mB,\mA)$. 
In what follows, $\gamma:\mA\otimes\mB\to\mB\otimes\mA$ will be used to denote the swap isomorphism. 
Every finite-dimensional $C^*$-algebra $\mA=\bigoplus_{x\in X}\matr_{m_{x}}(\C)$ has a unique positive functional $\tr:\mA\to\C$, called the \emph{trace}, such that $\tr\circ\gamma=\tr$ and $\tr(\mathds{1}_{m_{x}})=m_{x}$ for all $x\in X$. Its evaluation on an element of the form $\bigoplus_{x\in X}A_{x}$ is given by $\sum_{x\in X}\tr(A_{x})$ in terms of the usual trace on matrices. Every functional $\omega:\mA\to\C$ is given by $\tr(\rho\;\cdot\;)$, where $\rho:=\omega^{*}(1)^{\dag}\in\mA$ is the \emph{density} associated with $\omega$. The functional $\omega$ is $\dag$-preserving if and only if $\rho$ is self-adjoint, and if $\omega$ is positive and unital, then $\omega$ is referred to as a \emph{state}.
\end{notation}

\bd
\label{defn:channelstate}
Let $F:\mB\to\mA$ be a linear map. 
The \define{bloom} of $F$ is the linear map $\bloom_{F}:\mA\otimes\mB\to\mA$ given by $\bloom_{F}:=\mu_{\mA}\circ(\id_{\mA}\otimes F)$. 
The \define{swapped bloom} of $F$ is the linear map $\Lbloom{F}:\mA\otimes\mB\to\mA$ given by $\Lbloom{F}:=\mu_{\mA}\circ(F\otimes\id_{\mA})\circ\gamma$. 
The \define{channel state} associated with $F$ is the functional on $\mA\otimes\mB$ given by 
\[
\Cs[F]:=\tr\circ\bloom_{F}\equiv\tr\circ\mu_{\mA}\circ(\id\otimes F).
\]
The 
\define{channel density} associated with $F$ 
is the element of $\mA\otimes\mB$ given by 
\[
\Cd[F]:=\bloom_{F}^*(1_{\mA})^{\dag}\equiv(\id\otimes F^*)\big(\mu_{\mA}^*(1_{\mA})\big)^{\dag}
\]
and is the unique element of $\mA\otimes\mB$ that satisfies $\Cs[F]=\tr(\Cd[F]\;\cdot\;)$. 
The channel state provides a linear isomorphism $\Cs:\hom(\mB,\mA)\to\hom(\mA\otimes\mB,\C)$ sending $F$ to $\Cs[F]$ and the channel density provides a linear isomorphism $\Cd:\hom(\mB,\mA)\to\mA\otimes\mB$ sending $F$ to $\Cd[F]$. These isomorphisms will both be referred to as the \define{Choi--Jamio{\l}kowski isomorphism}~\cite{Ja72}. For additional clarity, these may also be written as $\Cs_{\mA,\mB}$ and $\Cd_{\mA,\mB}$. 
\ed

\br
In~\cite{HHPBS17}, the element $\Cd[F]$ is called the `channel state' associated with $F$ (this is because if $F$ is positive, then $\Cd[F]^{\dag}=\Cd[F]$---see Lemma~\ref{lem:channelstatesa}). 
We have chosen to call this the channel density to allow for simpler generalizations to the $C^*$-algebraic setting of hybrid classical/quantum systems. We should also point out that the state/density terminology is abusive for two reasons. First, the trace of $\Cd[F]$ is not unity, i.e., the channel state $\Cs[F]$ is not unital, even when $F$ is unital. Second, and more importantly, $\Cd[F]$ (and likewise $\Cs[F]$) need not be positive, even if $F$ is completely positive.%
\footnote{Note that $\Cd[F]$ is not the Choi matrix of $F$, the latter of which \emph{is} positive if and only if $F$ is completely positive. See the end of Remark~\ref{rmk:cupsandcaps} for more clarification.}
Our terminology is chosen to be somewhat consistent with that of~\cite{HHPBS17} (though what we call `density' is what~\cite{HHPBS17} calls a `state'). Nevertheless, the bloom and swapped bloom of $F$ are both unital if $F$ is unital.
Moreover, given a functional $\omega:\mathcal{A}\to\C$, the density $\mathscr{D}[\omega]$ is viewed as an element of $\mathcal{A}$ rather than $\C\otimes\mathcal{A}$, and satisfies $\omega=\tr(\mathscr{D}[\omega]\;\cdot\;)$. 
\er

\br
\label{rmk:cupsandcaps}
The channel state could have equivalently been defined in terms of the swapped bloom as $\Cs[F]=\tr\circ\Lbloom{F}$. The fact that these two are equal is a consequence of the properties of the trace. Furthermore, it will often be useful to depict this via string diagrams as
\[
\tr\circ\bloom_{F}
=\vcenter{\hbox{
\begin{tikzpicture}[font=\small]
\node[trace] (p) at (0,0) {};
\node[copier] (copier) at (0,0.3) {};
\node[arrow box] (g) at (0.5,0.95) {$F$};
\coordinate (X) at (-0.5,1.5);
\coordinate (Y) at (0.5,1.5);
\draw (p) to (copier);
\draw (copier) to[out=150,in=-90] (X);
\draw (copier) to[out=15,in=-90] (g);
\draw (g) to (Y);
\end{tikzpicture}}}
\qquad
\text{ and }
\qquad
\tr\circ\Lbloom{F}
=
\vcenter{\hbox{
\begin{tikzpicture}[font=\small]
\node[trace] (p) at (0,-0.5) {};
\node[copier] (copier) at (0,-0.15) {};
\coordinate (R) at (0.5,0.3) {};
\coordinate (Ls) at (-0.5,1.6) {};
\coordinate (Rs) at (0.5,1.6) {};
\coordinate (s2) at (-0.5,1.7) {};
\coordinate (s3) at (0.5,1.7) {};
\coordinate (g) at (0.5,0.7) {};
\node[arrow box] (h) at (-0.5,0.5) {$F$};
%
\draw (p) to (copier);
\draw (copier) to [out=15,in=-90] (g);
\draw (g) to [out=90,in=-90] (Ls);
\draw (h) to [out=90,in=-90] (Rs);
\draw (Ls) to (s2);
\draw (Rs) to (s3);
\draw (copier) to[out=165,in=-90] (h);
\end{tikzpicture}}}
\quad
,
\]
where $\vcenter{\hbox{\begin{tikzpicture}[font=\small]
\node[trace] (p) at (0,-0.35) {}; \draw (p) to (0,0); \end{tikzpicture}}}$ denotes the (un-normalized) trace, which is the Hilbert--Schmidt adjoint of $\shriek_{\mA}\equiv \vcenter{\hbox{\begin{tikzpicture}[font=\small]
\node[discarder] (p) at (0,0.35) {}; \draw (p) to (0,0); \end{tikzpicture}}}$.
Note that if $\mA=\bigoplus_{x\in X}\matr_{m_{x}}(\C)$, then 
\[
\vcenter{\hbox{
\begin{tikzpicture}[font=\small]
\node[discarder] (q) at (0,0.3) {};
\node[trace] (p) at (0,-0.3) {};
\draw (p) to (q);
\end{tikzpicture}}}
\quad
=
\quad
\sum_{x\in X}m_{x}
\]
gives the dimension of the underlying Hilbert space for $\mA$. Furthermore, if we set
\[
\vcenter{\hbox{
\begin{tikzpicture}[font=\small]
\coordinate (copier) at (0,0.3) {};
\coordinate (X) at (-0.5,0.7);
\coordinate (Y) at (0.5,0.7);
\draw (copier) to[out=180,in=-90] (X);
\draw (copier) to[out=0,in=-90] (Y);
\end{tikzpicture}}}
\quad:=\quad
\vcenter{\hbox{
\begin{tikzpicture}[font=\small]
\node[trace] (p) at (0,0) {};
\node[copier] (copier) at (0,0.3) {};
\coordinate (X) at (-0.5,0.7);
\coordinate (Y) at (0.5,0.7);
\draw (p) to (copier);
\draw (copier) to[out=165,in=-90] (X);
\draw (copier) to[out=15,in=-90] (Y);
\end{tikzpicture}}}
\qquad\text{ and }\qquad
\vcenter{\hbox{
\begin{tikzpicture}[font=\small,yscale=-1]
\coordinate (copier) at (0,0.3) {};
\coordinate (X) at (-0.5,0.7);
\coordinate (Y) at (0.5,0.7);
\draw (copier) to[out=180,in=-90] (X);
\draw (copier) to[out=0,in=-90] (Y);
\end{tikzpicture}}}
\quad:=\quad
\vcenter{\hbox{
\begin{tikzpicture}[font=\small,yscale=-1]
\node[discarder] (p) at (0,0) {};
\node[copier] (copier) at (0,0.3) {};
\coordinate (X) at (-0.5,0.7);
\coordinate (Y) at (0.5,0.7);
\draw (p) to (copier);
\draw (copier) to[out=165,in=-90] (X);
\draw (copier) to[out=15,in=-90] (Y);
\end{tikzpicture}}}
\quad,
\]
where $\vcenter{\hbox{
\begin{tikzpicture}[font=\small,yscale=-1]
\coordinate (p) at (0,0) {};
\node[copier] (copier) at (0,0.3) {};
\coordinate (X) at (-0.3,0.55);
\coordinate (Y) at (0.3,0.55);
\draw (p) to (copier);
\draw (copier) to[out=165,in=-90] (X);
\draw (copier) to[out=15,in=-90] (Y);
\end{tikzpicture}}}$ is the Hilbert--Schmidt adjoint of $\vcenter{\hbox{
\begin{tikzpicture}[font=\small]
\coordinate (p) at (0,0) {};
\node[copier] (copier) at (0,0.3) {};
\coordinate (X) at (-0.3,0.55);
\coordinate (Y) at (0.3,0.55);
\draw (p) to (copier);
\draw (copier) to[out=165,in=-90] (X);
\draw (copier) to[out=15,in=-90] (Y);
\end{tikzpicture}}}$, 
then the Choi--Jamio{\l}kowski isomorphism is seen to be an instance of the zig-zag identities from categorical quantum mechanics~\cites{CoKi17,HeVi19}. Note, however, that the cups 
$
\vcenter{\hbox{
\begin{tikzpicture}[font=\small,scale=0.5]
\coordinate (copier) at (0,0.3) {};
\coordinate (X) at (-0.5,0.7);
\coordinate (Y) at (0.5,0.7);
\draw (copier) to[out=180,in=-90] (X);
\draw (copier) to[out=0,in=-90] (Y);
\end{tikzpicture}}}
$
and caps 
$
\vcenter{\hbox{
\begin{tikzpicture}[font=\small,yscale=-1,scale=0.5]
\coordinate (copier) at (0,0.3) {};
\coordinate (X) at (-0.5,0.7);
\coordinate (Y) at (0.5,0.7);
\draw (copier) to[out=180,in=-90] (X);
\draw (copier) to[out=0,in=-90] (Y);
\end{tikzpicture}}}
$
here are not the usual ones from~\cites{CoKi17,HeVi19} involving maximally entangled states. This is because our cup takes a pure tensor $A\otimes B\in\matr_{m}(\C)\otimes\matr_{n}(\C)$ to $\tr(AB)$, whereas the cup defined using the un-normalized maximally entangled state sends $A\otimes B$ to $\tr(A^{T}B)\equiv\tr(AB^{T})$, where ${}^{T}$ denotes the transpose with respect to the basis chosen $\{|i\>\}$ that is used to define the maximally entangled vector $\sum_{i}|i\>\otimes|i\>$. Our basis-independent version is used for the Jamio{\l}kowski version of the Choi--Jamio{\l}kowski isomorphism~\cite{Ja72}, while the maximally entangled state is used to define the Choi version of the Choi--Jamio{\l}kowski isomorphism~\cite{Ch75}. In this paper, we exclusively use the Jamio{\l}kowski version.
\er

\bx[The channel density in the setting of matrix algebras]
\label{ex:channeldensitymatrixalgebra}
Let $F:\VB\to \VA$ be a completely positive unital map with $\mA=\matr_{m}(\C)$ and $\mB=\matr_{n}(\C)$. It then follows that the channel density defined in Definition~\ref{defn:channelstate} is the matrix given by (cf.\ Lemma~\ref{lem:channelstatesa})
\begin{equation}
\label{eq:DFmatrix}
\Cd[F]
=\sum_{i,j}^{m}E_{ij}^{(m)}\otimes F^*(E_{ji}^{(m)}),
\end{equation}
which coincides with the `channel state' $E_{\VB|\VA}$ associated with the completely positive trace-preserving map $F^*$ as defined in ~\cite[Section~2(a)]{HHPBS17}.%
\footnote{There is a small typo in the formula from~\cite[Section~2(a)]{HHPBS17} since the matrix $E_{\VB|\VA}$ should be an element of $\mA\otimes\mB$ and not $\mB\otimes\mA$.}
In particular, 
\[
\Cd[\id_{\mA}]=\mu_{\mA}^*(\mathds{1}_{m})=\sum_{i,j}^{m}E_{ij}^{(m)}\otimes E_{ji}^{(m)}.
\]
\ex

\bd
\label{defn:classicalmodel}
A pair $(F,\omega)$, where $F:\mB\to\mA$ and $\omega:\mA\to\C$ are linear, is \define{effectively classical}, or \define{has a classical model}, iff there exist commutative $C^*$-subalgebras $\mA_{\mathrm{cl}}\subseteq\mA$ and $\mB_{\mathrm{cl}}\subseteq\mB$, a linear map $F_{\mathrm{cl}}:\mB_{\mathrm{cl}}\to\mA_{\mathrm{cl}}$, and conditional expectations%
\footnote{This means $E_{\mA}$ and $E_{\mB}$ are positive unital and satisfy $E_{\mA}\circ j_{\mA}=\id_{\mA_{\mathrm{cl}}}$ and $E_{\mB}\circ j_{\mB}=\id_{\mB_{\mathrm{cl}}}$. See~\cite{GPRR21} for further properties.}
$E_{\mA}:\mA\to\mA_{\mathrm{cl}}$ and $E_{\mB}:\mB\to\mB_{\mathrm{cl}}$ such that 
\[
\omega_{\restriction}\circ E_{\mA}=\omega, 
\qquad
(\omega\circ F)_{\restriction}\circ E_{\mB}=\omega\circ F, 
\qquad
\text{ and }
\qquad
F=j_{\mA}\circ F_{\mathrm{cl}}\circ E_{\mB}, 
\]
where $j_{\mA}:\mA_{\mathrm{cl}}\to\mA$ is the inclusion, and where $\restriction$ is used to denote the restriction, i.e., $\omega_{\restriction}:=\omega\circ j_{\mA}$ and $(\omega\circ F)_{\restriction}:=\omega\circ F\circ j_{\mB}$. 
\ed
 
The first two equations in Definition~\ref{defn:classicalmodel} say that the conditional expectations are state-preserving, i.e., they are particular disintegrations in the terminology of~\cite{GPRR21}. The last condition is more easily visualized as the commutative diagram 
\[
\xy0;/r.25pc/:
(-10,7.5)*+{\mB_{\mathrm{cl}}}="1";
(10,7.5)*+{\mA_{\mathrm{cl}}}="2";
(-10,-7.5)*+{\mB}="3";
(10,-7.5)*+{\mA}="4";
{\ar"3";"1"^{E_{\mB}}};
{\ar"1";"2"^{F_{\mathrm{cl}}}};
{\ar"2";"4"^{j_{\mA}}};
{\ar"3";"4"_{F}};
\endxy
\]
and encapsulates the fact that the quantum dynamics factors through a classical system. 
The conditional expectation condition also guarantees that the composite of effectively classical channels factors through the composite of the underlying classical channels. This is not relevant for the main statement of our theorem and is therefore addressed later in Proposition~\ref{prop:composingclassicalmodels}. 
The motivation for Definition~\ref{defn:classicalmodel} comes from the argument often provided in the physics literature that density matrices and channels that can be `simultaneously diagonalized' are classical. The precise statement is given in the next proposition, the proof of which is given in the next section. 

\bn
\label{prop:quantumclassical}
In the notation of Definition~\ref{defn:classicalmodel}, let $\mA=\matr_{m}(\C)$ and $\mB=\matr_{n}(\C)$, and suppose $\omega$ and $F$ are $\dag$-preserving. 
Write $\rho:=\mathscr{D}[\omega]$ and $\theta:=\mathscr{D}[\omega\circ F]$.
Then, a classical model for $(F,\omega)$ exists if and only if there exist orthonormal bases $\{e_{i}\}$ for $\C^{m}$ and $\{\epsilon_{k}\}$ for $\C^{n}$ such that $\rho$ and $\theta$ are diagonal in these bases and the channel density $\Cd[F]$ associated with $F$ is diagonal%
\footnote{It suffices to assume that $\rho$ and $\Cd[F]$ are diagonal in these bases since $\theta=F^*(\rho)$ will be diagonal as a consequence.}
in the basis $\{e_{i}\otimes \epsilon_{k}\}$. Furthermore, if these equivalent conditions hold, then $[\Cd[F],\rho\otimes1_{\mB}]=0$.
\en

\bd
\label{defn:statesovertime}
Let $\star$ be a family of functions that assigns to any pair of finite-dimensional $C^*$-algebras $\mA$ and $\mB$ a function
$\star:\hom(\mB,\mA)\times\hom(\mA,\C)\to\hom(\mA\otimes\mB,\C)$ taking any linear map $F:\mB\to\mA$ and linear functional $\omega$ on $\mA$ to a functional $F\star\omega$ on $\mA\otimes\mB$ and satisfying the following conditions.
\begin{enumerate}[(a)]
\itemsep0em
\item
(Hermiticity and unitality) If $(F,\omega)$ is in $\sahom(\mB,\mA)\times\sahom(\mA,\C)$, then $F\star\omega$ is in $\sahom(\mA\otimes\mB,\C)$. If $F$ and $\omega$ are both unital, then $F\star\omega$ is unital. 
\item
(Preservation of probabilistic mixtures/convex bi-linearity)
Given any $\lambda\in[0,1]$ together with maps $F,G\in\hom(\mB,\mA)$ and $\omega,\xi\in\hom(\mA,\C)$, the equalities
\[
\Big(\lambda F+(1-\lambda)G\Big)\star\omega=\lambda(F\star\omega)+(1-\lambda)(G\star\omega)
\]
and
\[
F\star\Big(\lambda\omega+(1-\lambda)\xi\Big)=\lambda(F\star\omega)+(1-\lambda)(F\star\xi)
\]
hold.
\item
(Preservation of classical limit) 
If the pair $(F,\omega)$ is effectively classical, then 
$F\star\omega=\omega\circ\bloom_{F}$.
\item
(Preservation of marginal states) The initial and final functionals are recovered from the joint functional in the sense that 
\[
(F\star\omega)\circ i_{\mA}=\omega
\qquad\text{ and }\qquad
(F\star\omega)\circ i_{\mB}=\omega\circ F
\]
for all unital $F$.
\item
(Compositionality/associativity)
Given a composable pair of unital 
maps $\mC\xrightarrow{G}\mB\xrightarrow{F}\mA$ and a unital 
functional $\omega$ on $\mA$, 
\[
(\shriek_{\mA}\otimes G)\star(F\star\omega)
=\Big(\Cs_{\mA,\mB\otimes\mC}^{-1}\Big[(\shriek_{\mA}\otimes G)\star\big(\Cs_{\mA,\mB}[F]\big)\Big]\Big)\star\omega. 
\]
\end{enumerate}
Such a family is called a \define{states over time} function.
\ed

The explanation for why the compositionality/associativity formula looks so complicated, but is in fact rather straightforward, will be given in Remark~\ref{rmk:associativity}. In short, it follows from the two natural ways of pairing the construction of states over two successive times and only looks complicated due to the natural isomorphisms coming from the Choi--Jamio{\l}kowski isomorphism. Secondly, although our preservation of the classical limit axiom is expressed differently than in~\cite{HHPBS17}, it is equivalent to it by Proposition~\ref{prop:quantumclassical} on the domains for which a family of states over time function is defined.

\br
\label{rmk:statesovertimeterminology}
The terminology `a family of \emph{states} over time' is a bit abusive because we are only requiring $F\star\omega$ to be $\dag$-preserving and unital in axiom (a) of Definition~\ref{defn:statesovertime}, rather than positive if $F$ and $\omega$ are positive. Note that this is the same restriction imposed in~\cite{HHPBS17}. In other words, there are situations where one might begin with a positive (even completely positive) unital map together with a state and end up with a joint functional that is \emph{not} a state, i.e., it is not necessarily positive. It is an open question whether one can obtain a genuine family of states over time where all maps remain positive under some operation satisfying similar, perhaps slightly weakened, axioms (see Section~\ref{sec:discussion} for more details). 
\er

\br
\label{rmk:hermitianextend}
Rather than requiring a family of states over time function to be defined as a family of functions of the form $\hom(\mB,\mA)\times\hom(\mA,\C)\to\hom(\mA\otimes\mB,\C)$, we could have required it to be a family of functions of the form $\sahom(\mB,\mA)\times\sahom(\mA,\C)\to\sahom(\mA\otimes\mB,\C)$ so that Hermiticity is in the very definition of the family. However, one can see that by arguments completely analogous to those in the proof of~\cite[Lemma~4.3]{HHPBS17}, any such function uniquely extends to a function $\hom(\mB,\mA)\times\hom(\mA,\C)\to\hom(\mA\otimes\mB,\C)$ satisfying the same properties, in fact complex bi-linearity. This is achieved by splitting an arbitrary morphism $F:\mB\to\mA$ into its Hermitian and anti-Hermitian parts via $F=\frac{1}{2}\left(F+\dag\circ F\circ\dag\right)+\frac{1}{2}\left(F-\dag\circ F\circ\dag\right)$. Thus, we lose no generality in defining a family of states over time function on all linear maps as opposed to the subspace of $\dag$-preserving ones.
\er

\br
\label{rmk:associativity}
The formula for associativity looks rather complicated because of the way in which we have formulated our definition by avoiding the usage of a binary operation and is one of the two reasons why we are able to bypass the no-go result of~\cite{HHPBS17} (see Remark~\ref{rmk:bypassHHPBS17} for more details). The formula comes from trying to pair the three different factors in the two possible ways, i.e., the diagram%
\[
{\small
\xy0;/r.25pc/:
(-48,20)*+{\hom(\mC,\mB)\times\hom(\mB,\mA)\times\hom(\mA,\C)}="1";
(48,20)*+{\hom(\mC,\mA\otimes\mB)\times\hom(\mA\otimes\mB,\C)\times\hom(\mA,\C)}="2";
(-48,7)*+{\hom(\mC,\mB)\times\hom(\mA\otimes\mB,\C)}="3";
(48,7)*+{\hom(\mA\otimes\mB\otimes\mC,\C)\times\hom(\mA,\C)}="4";
(-48,-7)*+{\hom(\mC,\mA\otimes\mB)\times\hom(\mA\otimes\mB,\C)}="5";
(48,-7)*+{\hom(\mB\otimes\mC,\mA)\times\hom(\mA,\C)}="6";
(0,-20)*+{\hom(\mA\otimes\mB\otimes\mC,\C)}="7";
{\ar"1";"2"^(0.46){(\shriek_{\mA}\otimes\;\cdot\;)\times\Cs_{\mA,\mB}\times\id}};
{\ar"1";"3"^{\id\times\star}};
{\ar"2";"4"_{\star\times\id}};
{\ar"3";"5"^{(\shriek_{\mA}\otimes\;\cdot\;)\times\id}};
{\ar"4";"6"_{\Cs_{\mA,\mB\otimes\mC}^{-1}\times\id}};
{\ar"5";"7"^{\star}};
{\ar"6";"7"_{\star}};
\endxy
}
\]
must commute. Note that the Choi--Jamio{\l}kowski isomorphisms $\Cs$ are used to transform channels into joint states, while the inverse transforms joint states back into channels. Furthermore, the inclusion $\shriek_{\mA}$ is used to guarantee that the domains and codomains match so that the $\star$ operation can be applied. In other words, ignoring these canonical maps, one sees this as associativity of $\star$. Interestingly, when we extend such a $\star$ function to include channels in its second argument in Section~\ref{sec:channelsovertime}, we will find that the associated formulation of associativity takes a much simpler form, as it will no longer be necessary to use the Choi--Jamio{\l}kowski isomorphism in its description.
\er

Our main result is the following theorem, which we will prove in Section~\ref{sec:proofs}.

\bt
\label{thm:statesovertime}
The function sending $F:\mB\to\mA$ and $\omega:\mA\to\C$ to 
\[
F\star\omega:=\frac{1}{2}\Big(\omega\circ\bloom_{F}+\omega\circ\Lbloom{F}\Big)
\equiv\frac{1}{2}\left(
\vcenter{\hbox{
\begin{tikzpicture}[font=\small]
\node[state] (p) at (0,0) {$\omega$};
\node[copier] (copier) at (0,0.3) {};
\node[arrow box] (g) at (0.5,0.95) {$F$};
\coordinate (X) at (-0.5,1.5);
\coordinate (Y) at (0.5,1.5);
\draw (p) to (copier);
\draw (copier) to[out=150,in=-90] (X);
\draw (copier) to[out=15,in=-90] (g);
\draw (g) to (Y);
\end{tikzpicture}}}
\;
+
\;
\vcenter{\hbox{
\begin{tikzpicture}[font=\small]
\node[state] (p) at (0,-0.5) {$\omega$};
\node[copier] (copier) at (0,-0.15) {};
\coordinate (R) at (0.5,0.3) {};
\coordinate (Ls) at (-0.5,1.6) {};
\coordinate (Rs) at (0.5,1.6) {};
\coordinate (s2) at (-0.5,1.7) {};
\coordinate (s3) at (0.5,1.7) {};
\coordinate (g) at (0.5,0.7) {};
\node[arrow box] (h) at (-0.5,0.5) {$F$};
%
\draw (p) to (copier);
\draw (copier) to [out=15,in=-90] (g);
\draw (g) to [out=90,in=-90] (Ls);
\draw (h) to [out=90,in=-90] (Rs);
\draw (Ls) to (s2);
\draw (Rs) to (s3);
\draw (copier) to[out=165,in=-90] (h);
\end{tikzpicture}}}
\right)
\]
is a states over time function.
\et

\br[States over time function between matrix algebras in the Schr\"odinger picture, i.e., the setting of completely positive trace-preserving (CPTP) maps]
Given a completely positive unital map $F:\VB\to \VA$ with $\VB=\M_n(\C)$, $\VA=\M_m(\C)$ and a density matrix $\rho=\mathscr{D}[\omega]$ associated with a state $\omega:\VA\to \C$, it follows from Lemma~\ref{lem:jordansa} that the density $\mathscr{D}[F\star \omega]$ is given by the Jordan product of $\mathscr{D}[F]$ and $\rho\otimes \mathds{1}_{n}$, i.e.,
\[
\mathscr{D}[F\star \omega]=\frac{1}{2}\big(\mathscr{D}[F](\rho\otimes \mathds{1}_{n})+(\rho\otimes \mathds{1}_{n})\mathscr{D}[F]\big).
\]
Together with this and the expression~(\ref{eq:DFmatrix}) for $\Cd[F]$ 
from Example~\ref{ex:channeldensitymatrixalgebra}, the density $\mathscr{D}[F\star \omega]$ may be explicitly expressed
in terms of the CPTP map $F^*$ and the density matrix $\rho=\mathscr{D}[\omega]\in \VA$. As such, if we associate with every CPTP map $\varphi:\VA\to \VB$ the matrix $\mathscr{C}[\varphi]\in \VA\otimes \VB$ given by%
\footnote{The matrix $\mathscr{C}[\varphi]$ is denoted by $E_{\VB|\VA}$ in \cite{HHPBS17}.}
\[
\mathscr{C}[\varphi]=\sum_{i,j}^{m}E_{ij}^{(m)}\otimes \varphi(E_{ji}^{(m)}),
\]
then one may reformulate the states over time function as defined in Theorem~\ref{thm:statesovertime} so that when restricted to CPTP maps between matrix algebras as a function $\star:\CPTP(\mA,\mB)\times \mA\to \VA\otimes \VB$, it is given by the Jordan product, namely
\[
\varphi\star \rho=\frac{1}{2}\big(\mathscr{C}[\varphi](\rho\otimes \mathds{1}_{n})+(\rho\otimes \mathds{1}_{n})\mathscr{C}[\varphi]\big),
\]
where $\CPTP(\mA,\mB)$ denotes the set of CPTP maps from $\VA=\M_m(\C)$ to $\VB=\M_n(\C)$. 

Moreover, one may show that such a states over time function satisfies the appropriate translations of axioms (a)--(e) in Definition~\ref{defn:statesovertime} to the setting of CPTP maps between matrix algebras. In particular, if $\VA\overset{\varphi}\to \VB\overset{\psi}\to \mathcal{C}$ is a pair of composable CPTP maps between matrix algebras, and if we let $\widetilde{\mathscr{C}}[\psi]={1}_{\VA}\otimes \mathscr{C}[\psi]$, $\widetilde{\mathscr{C}}[\varphi]=\mathscr{C}[\varphi]\otimes {1}_{\VC}$, and  $\widetilde{\rho}=\rho\otimes {1}_{\VB}\otimes {1}_{\VC}$, then the associativity axiom in this context translates to
\[
\widetilde{\mathscr{C}}[\psi]\star \left(\widetilde{\mathscr{C}}[\varphi]\star \widetilde{\rho}\right)=\left(\widetilde{\mathscr{C}}[\psi]\star \widetilde{\mathscr{C}}[\varphi]\right)\star \widetilde{\rho},
\]
which indeed holds since $[\widetilde{\mathscr{C}}[\psi],\widetilde{\rho}]=0$.  
\er

\bx[The Einstein--Podolsky--Rosen (EPR) state as a state over time]
Let $\mA=\mB=\matr_{2}(\C)$, let $\omega=\frac{1}{2}\tr$, and let $F:\mB\to\mA$ be the map given by
\[
F(B)=\begin{bmatrix}0&1\\-1&0\end{bmatrix}B^{T}\begin{bmatrix}0&-1\\1&0\end{bmatrix}. 
\]
Then $F$ is positive and unital, but not completely positive. Nevertheless, the associated state over time $F\star\omega$ via the function from Theorem~\ref{thm:statesovertime} has density given by
\[
\mathscr{D}[F\star\omega]=\frac{1}{2}\begin{bmatrix}0&0&0&0\\0&1&-1&0\\0&-1&1&0\\0&0&0&0\end{bmatrix},
\]
which corresponds to the EPR state~\cites{PaQPL21,EPR,Bo51}. 
In this way, the EPR state, which is an entangled state on $\mA\otimes\mB$ at \emph{a single time} can also be viewed as a state \emph{over time} using the initial maximally mixed state $\omega$ and the positive (but not completely positive) inference map $F$, together with our prescription from Theorem~\ref{thm:statesovertime}.
\ex

\br[Incompatibility and temporal correlations in the setting of matrix algebras]
\label{rmk:incompatibility}
The Jordan product used in our construction of a states over time function plays an important role in the resource theory of incompatibility of channels (and in particular, measurements)~\cite{GPS21}. In addition, in the special case of qubits and a single quantum channel describing evolution, our construction contains information about the temporal correlations between the initial and final states. In particular, the works~\cites{FJV15,ZPTGVF18} argue that a lack of positivity of the associated states over time provides quantitative evidence for such temporal correlations. 

On the other hand, one can guarantee positivity of our state over time by perturbing a channel and an initial state with sufficient noise. For example, consider the case of an initial density matrix of the form $\rho=\left[\begin{smallmatrix}p&0\\0&1-p\end{smallmatrix}\right]$ with $p\in[0,1]$ and $F=\id_{\matr_{2}(\C)}$. If we now add noise to our initial state $\omega=\tr(\rho\;\cdot\;)$ and our map $F$ in the form 
\[
\omega\mapsto(1-\epsilon)\omega+\frac{\epsilon}{2}\tr
\quad\text{ and }\quad
F\mapsto(1-\delta)F+\frac{\delta}{2}\left(\shriek_{\matr_{2}(\C)}\circ\tr\right),
\]
where $\epsilon,\delta\in[0,1],$ 
then the resulting state over time becomes
\[
F\star\omega\mapsto(1-\delta)(1-\epsilon)(F\star\omega)+(1-\delta)\epsilon(\omega\circ\bloom_{F})+\frac{\delta(1-\epsilon)}{2}(\omega\otimes\tr)+\frac{\delta\epsilon}{4}(\tr\otimes\tr),
\]
which follows from linearity of our states over time function in \emph{both} variables and the symmetric property of the trace. The associated density therefore becomes 
\[
\mathscr{D}[F\star\omega]\mapsto\begin{bmatrix}\left(1-\frac{\delta}{2}\right)\left(\epsilon^{\perp}p+\frac{\epsilon}{2}\right)&0&0&0\\0&\frac{\delta}{2}\left(\epsilon^{\perp}p+\frac{\epsilon}{2}\right)&\frac{1-\delta}{2}&0\\0&\frac{1-\delta}{2}&\frac{\delta}{2}\left(\epsilon^{\perp}p^{\perp}+\frac{\epsilon}{2}\right)&0\\0&0&0&\left(1-\frac{\delta}{2}\right)\left(\epsilon^{\perp}p^{\perp}+\frac{\epsilon}{2}\right)\end{bmatrix},
\]
where we have temporarily introduced the notation $\epsilon^{\perp}:=1-\epsilon$ and $p^{\perp}:=1-p$. The only eigenvalue that becomes negative for some values of the parameters is given by 
\[
\lambda(p,\delta,\epsilon):=\frac{1}{4}\left(\delta-\sqrt{4-8\delta+\delta^2(5-4p(1-\epsilon)^2+4p^2(1-\epsilon)^2-2(1-\epsilon)\epsilon)}\right).
\]
\begin{figure}
  \centering
    \includegraphics[width=0.4\textwidth]{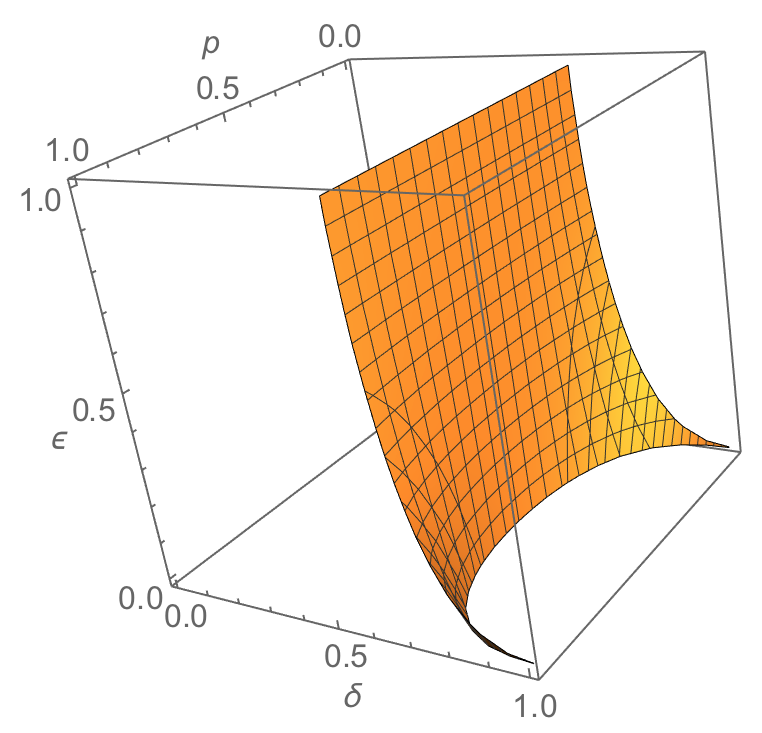}
  \caption{By adding enough noise to the density matrix $\rho=\mathrm{diag}(p,1-p)$ and the channel $F=\id_{\matr_{2}(\C)}$, described by the parameters $\delta$ and $\epsilon$, the state over time eventually becomes a genuine state. The region to the right of the surface (for large enough values of $\delta$ and $\epsilon$) depicts where positivity is achieved.}
  \label{fig:PAN}
\end{figure}

Figure~\ref{fig:PAN} shows a plot of the surface where this eigenvalue vanishes. This splits the cube $[0,1]^{3}\subseteq\R^{3}$, describing the three parameters $p,\delta,$ and $\epsilon$, into two disjoint regions, where on one side the associated density has a negative eigenvalue and therefore cannot be interpreted as a density matrix, while on the other side the associated density has only non-negative eigenvalues and hence can be interpreted as a genuine density matrix. 
\er

\begin{figure}
    \centering
    \begin{subfigure}[b]{0.47\textwidth}
        \includegraphics[width=\textwidth]{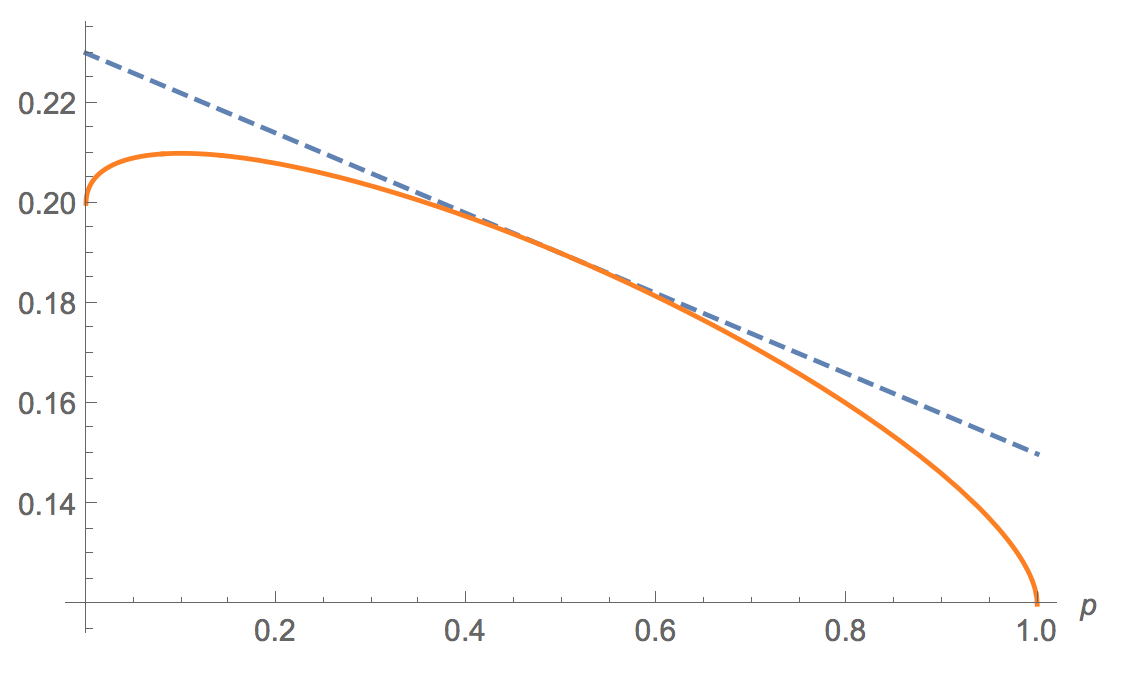}
        \caption{Here,  $M=\left[\begin{smallmatrix}0.4&-0.2+0.2i\\-0.2-0.2i&0.8\end{smallmatrix}\right]$ and $N=\left[\begin{smallmatrix}0.6&-0.4-0.1i\\-0.4+0.1i&0.5\end{smallmatrix}\right]$. In this case, the tangent line is always above zero.}
    \end{subfigure}
    \qquad
    \begin{subfigure}[b]{0.47\textwidth}
        \includegraphics[width=\textwidth]{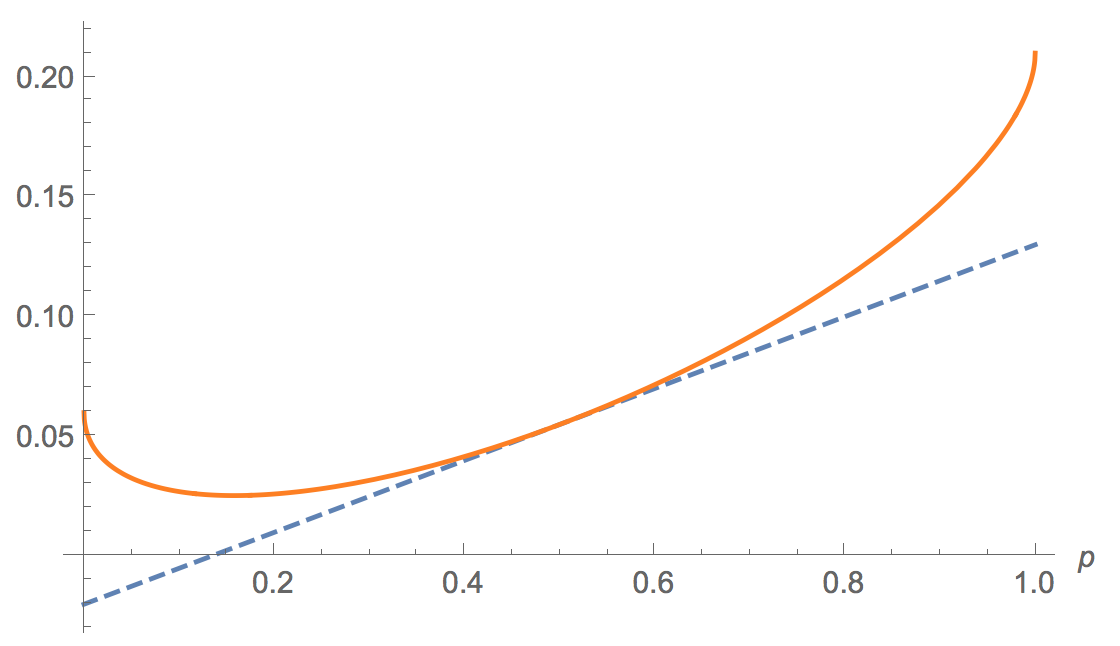}
        \caption{Here, $M=\left[\begin{smallmatrix}0.6&-0.4\\-0.4&0.3\end{smallmatrix}\right]$ and $N=\left[\begin{smallmatrix}0.7&0.4\\0.4&0.4\end{smallmatrix}\right]$. In this case, there exist values of $p$ for which the tangent line goes below zero. }
    \end{subfigure}
    \caption{Two examples of probabilities associated with specified measurement operators (effects) $M$ and $N$ for Alice and Bob, respectively, using the Leifer--Spekkens (LS) state over time (solid orange line) versus our prescription (dashed blue line) from Theorem~\ref{thm:statesovertime}. The horizontal axis corresponds to the initial state $\mathrm{diag}(p,1-p)$, while the vertical axis corresponds to the associated probabilities. It is always the case that the prescription from Theorem~\ref{thm:statesovertime} provides the tangent line to the LS prescription.}
    \label{fig:LStangentFP}
\end{figure}

\br[Comparison with Leifer and Spekkens' State Over Time]
\label{rmk:operational}
Let $F:\VB\to \VA$ be a positive unital map with $\VA=\M_m(\C)$ and $\VB=\M_n(\C)$, and let $\omega:\VA\to \C$ be a state. In ~\cites{Le07,LeSp13}, Leifer and Spekkens define an associated state over time $F\star_{\mathrm{LS}}\omega$ whose associated density $\mathscr{D}\left[F\star_{\mathrm{LS}}\omega\right]$ is given by
\[
\mathscr{D}\left[F\star_{\mathrm{LS}}\omega\right]=(\sqrt{\rho}\otimes \mathds{1})\mathscr{D}[F](\sqrt{\rho}\otimes \mathds{1}),
\]
where $\rho=\mathscr{D}[\omega]$. Given the extra data of positive operator-valued measures $\C^{X}\xrightarrow{M}\mA$ and  $\C^{Y}\xrightarrow{N}\mB$, Leifer shows in \cite{Le06} that one may associate a probability distribution $\mathbb{P}(x,y)$ on $X\times Y$ given by
\be \label{LFXPX17}
\mathbb{P}(x,y)=\left(F\star_{\mathrm{LS}}\omega\right)(M_x\otimes N_y),
\ee
which is interpreted as the probability $x$ and $y$ occur when making local measurements $M_{x}$ and $N_{y}$ on $\VA$ and $\VB$, respectively. It is then natural to question how the probabilities $\mathbb{P}(x,y)$ compare with the right-hand side of \eqref{LFXPX17} when $F\star_{\mathrm{LS}}\omega$ is replaced with our state over time $F\star\omega$ as defined in Theorem~\ref{thm:statesovertime}. What we find is that $\left(F\star \omega\right)(M_x\otimes N_y)$ provides the  \emph{linear} approximation to $\mathbb{P}(x,y)$ in a neighborhood of the maximally mixed state corresponding to the normalized identity matrix. Examples involving qubits and the family of initial states $\omega_p$ with $\mathscr{D}[\omega_p]=\left[\begin{smallmatrix}p&0\\0&1-p\end{smallmatrix}\right]$ for all $p\in[0,1]$ and $F=\id_{\matr_{2}(\C)}$ are depicted in Figure~\ref{fig:LStangentFP}. The general statement and its proof will appear elsewhere.
\er

\br
\label{rmk:bypassHHPBS17}
Theorem~\ref{thm:statesovertime} seems to contradict the results of~\cite{HHPBS17}. However, as mentioned in Remark~\ref{rmk:associativity}, there are two reasons for this. 
First, by using the Choi--Jamio{\l}kowski isomorphism to identify $\hom(\mB,\mA)$ with $\mA\otimes\mB$, the authors of~\cite{HHPBS17} demanded the five axioms of a state over time to be valid on the larger domain $(\mA\otimes\mB)\times (\mA\otimes\mB)$, rather than the domain 
\[
\hom(\mB,\mA)\times\hom(\mA,\C)\cong(\mA\otimes\mB)\times(\C\otimes\mA)\cong(\mA\otimes\mB)\times\mA,
\]
as we have.
The second reason, which will be explained in more detail in Section~\ref{sec:channelsovertime}, is that the extension of the classical limit axiom as formulated in~\cite{HHPBS17} to the larger domain is more stringent than our extension based on our definition of a classical model. By using only the data given (a channel together with a state), and by using less structure that is nevertheless sufficient to make sense of such a family of states over time function, we have been able to bypass the no-go theorem of~\cite{HHPBS17}. It should be pointed out that the authors of~\cite{HHPBS17} were well aware of the fact that the Jordan product on qubits is associative on the required matrices involved when they discussed the Fitzsimons--Jones--Vedral (FJV) construction~\cite{FJV15}. However, they nevertheless demanded a binary operation, so that the Jordan product is no longer associative, and this is what forced their no-go result. 
\er

\section{Proofs and relevant results}
\label{sec:proofs}

This section contains proofs of all statements made, including the proof of the main theorem. Some lemmas are included as separate statements and a proposition regarding the compositionality of classical models is given. 

\bprf
[Proof of Proposition~\ref{prop:quantumclassical}]
{\color{white}{you found me!}}

\noindent 
($\Leftarrow$)
Suppose there exist orthonormal bases $\{e_{i}\}$ and $\{\epsilon_{j}\}$ as in the assumptions of the reverse claim. Let $P_{i}\equiv|e_{i}\>\<e_{i}|$ and $Q_{k}\equiv|\epsilon_{k}\>\<\epsilon_{k}|$ be the corresponding one-dimensional projection operators. By assumption, there exist numbers $p_{i},q_{k},f_{ki}\in\R$ such that 
\[
\rho=\sum_{i=1}^{m}p_{i}P_{i}
,\qquad
\theta=\sum_{k=1}^{n}q_{k}Q_{k}
,\quad
\text{ and }\quad
\Cd[F]
=\sum_{i=1}^{m}\sum_{k=1}^{n}f_{ki}P_{i}\otimes Q_{k}. 
\]
By Example~\ref{ex:channeldensitymatrixalgebra}, the last equality for the channel density entails 
\[
F^*\big(|e_{j}\>\<e_{i}|\big)=\delta_{ij}F^*(P_{i})=\sum_{k=1}^{n}\delta_{ij}f_{ki}Q_{k}
\]
for all $i,j\in\{1,\dots,m\}$. 
Therefore, 
\[
F\big(|\epsilon_{k}\>\<\epsilon_{l}|\big)=\de_{kl}\sum_{i=1}^{m}\overline{f_{ki}}P_{i}
\]
for all $k,l\in\{1,\dots,n\}$ by the definition of the Hilbert--Schmidt adjoint (since the $f_{ki}$ are real, $\overline{f_{ki}}=f_{ki}$), since this definition satisfies 
\[
\tr\left(E_{ij}^{(m)}F\big(E_{kl}^{(n)}\big)\right)\equiv
\left\<E_{ji}^{(m)},F\big(E_{kl}^{(n)}\big)\right\>=\left\<F^*\big(E_{ji}^{(m)}\big),E_{kl}^{(n)}\right\>\equiv\tr\left(F^{*}\big(E_{ji}^{(m)}\big)^{\dag}E_{kl}^{(n)}\right)
\]
for all $i,j\in\{1,\dots,m\}$ and $k,l\in\{1,\dots,n\}$. 
From this, we can define the required conditional expectations and classical maps. 
First, we set $\mA_{\mathrm{cl}}:=\mathrm{span}_{i}\{P_{i}\}$ and $\mB_{\mathrm{cl}}:=\mathrm{span}_{k}\{Q_{k}\}$. These are commutative unital $C^*$-subalgebras of $\mA$ and $\mB$, respectively, due to the orthonormality and spanning assumptions on $\{e_{i}\}$ and $\{\epsilon_{k}\}$. Next, define $F_{\mathrm{cl}}:\mB_{\mathrm{cl}}\to\mA_{\mathrm{cl}}$ by specifying
\[
F_{\mathrm{cl}}(Q_{k}):=\sum_{i=1}^{m}\overline{f_{ki}}P_{i}
\]
for all $k$ and then extending linearly. The conditional expectations $E_{\mA}:\mA\to\mA_{\mathrm{cl}}$ and $E_{\mB}:\mB\to\mB_{\mathrm{cl}}$ are defined by 
\[
E_{\mA}(A):=\sum_{i=1}^{m}P_{i}AP_{i}
\qquad\text{ and }\qquad
E_{\mB}(B):=\sum_{k=1}^{n}Q_{k}BQ_{k}. 
\]
From this, the required conditions of a classical model are all readily checked. Indeed, one has
\[
(j_{\mA}\circ F_{\mathrm{cl}}\circ E_{\mB})\big(|\epsilon_{k}\>\<\epsilon_{l}|\big)
=(j_{\mA}\circ F_{\mathrm{cl}})(\delta_{kl}Q_{k})
=\delta_{kl}\sum_{i=1}^{m}\overline{f_{ki}}P_{i}
=F\big(|\epsilon_{k}\>\<\epsilon_{l}|\big),
\]
as needed.
The state-preserving condition for the conditional expectations is immediate. 

\noindent
($\Rightarrow$)
Suppose a classical model exists. Then by all the assumptions, there exist orthogonal projections (not necessarily rank 1) $\{P_{i}\}$ in $\mA$ and $\{Q_{k}\}$ in $\mB$ together with coefficients $\{p_{i}\}$ and $\{q_{k}\}$ such that 
\[
\mA_{\mathrm{cl}}=\mathrm{span}_{i}\{P_{i}\},
\quad
\mB_{\mathrm{cl}}=\mathrm{span}_{k}\{Q_{k}\},
\quad
\omega(P_{i})=p_{i},
\quad
\text{ and }
\quad
(\omega\circ F)(Q_{k})=q_{k}.
\] 
Since $E_{\mA}$ and $E_{\mB}$ are conditional expectations into commutative $C^*$-algebras, there exist positive functionals $\phi_{i}:\mA\to\C$ and $\psi_{k}:\mB\to\C$ supported on $P_{i}\mA P_{i}$ and $Q_{k}\mB Q_{k}$, respectively, such that 
\[
E_{\mA}(A)=\sum_{i}\phi_{i}(P_{i}A P_{i})P_{i}
\quad\text{ and }\quad
E_{\mB}(B)=\sum_{k}\psi_{k}(Q_{k}B Q_{k})Q_{k}.
\]
Such functionals are necessarily represented by positive matrices $\sigma_{i}\in P_{i}\mA P_{i}$ and $\tau_{k}\in Q_{k}\mB Q_{k}$ satisfying 
\[
\phi_{i}=\tr(\sigma_{i}\;\cdot\;)
\quad\text{ and }\quad \psi_{k}=\tr(\tau_{k}\;\cdot\;).
\]
By the assumption that the conditional expectations $E_{\mA}$ and $E_{\mB}$ are state-preserving, we conclude that $\rho$ and $\theta$ are (orthogonal) linear combinations of these matrices, namely
\[
\rho=\sum_{i}p_{i}\sigma_{i}
\quad
\text{ and }
\quad
\theta=\sum_{k}q_{k}\tau_{k}. 
\]
Since $\rho$ and $\theta$ are self-adjoint, $\sigma_{i}$ and $\tau_{k}$ are self-adjoint as well. As such, let $\{|e_{i \alpha_{i}}\>\}$ and $\{|\epsilon_{k \beta_{k}}\>\}$ be orthonormal bases diagonalizing the $\sigma_{i}$ and $\tau_{k}$, respectively. Thus, 
\[
\rho=\sum_{i}p_{i}\sum_{\alpha_{i}}s_{i \alpha_{i}}|e_{i \alpha_{i}}\>\<e_{i \alpha_{i}}|
\quad\text{ and }\quad
\theta=\sum_{k}q_{k}\sum_{\beta_{k}}t_{k \beta_{k}}|\epsilon_{k \beta_{k}}\>\<\epsilon_{k \beta_{k}}|
\]
have been diagonalized in terms of the real coefficients $\{s_{i\alpha_{i}}\}$ and $\{t_{k\beta_{k}}\}$. In particular, note that 
\[
P_{i}=\sum_{\alpha_{i}}|e_{i \alpha_{i}}\>\<e_{i \alpha_{i}}|
\quad\text{ and }\quad
Q_{k}=\sum_{\beta_{k}}|\epsilon_{k \beta_{k}}\>\<\epsilon_{k \beta_{k}}|
\]
provide orthogonal rank 1 decompositions of the projection operators describing the commutative subalgebras. 
Now, since $F_{\mathrm{cl}}:\mB_{\mathrm{cl}}\to\mA_{\mathrm{cl}}$ is linear and $\dag$-preserving, there exist numbers $f_{ki}\in\R$ such that 
\[
F_{\mathrm{cl}}(Q_{k})=\sum_{i}f_{ki}P_{i}
\]
for all $k$. By the assumption that $F=j_{\mA}\circ F_{\mathrm{cl}}\circ E_{\mB}$ together with all the consequences derived thus far, we find 
\[
\begin{split}
F\left(|\epsilon_{k \beta_{k}}\>\<\epsilon_{l \gamma_{l}}|\right)
&=(j_{\mA}\circ F_{\mathrm{cl}})\Big(\delta_{kl}\tr\big(\tau_{k}|\epsilon_{k \beta_{k}}\>\<\epsilon_{l \gamma_{l}}|\big)Q_{k}\Big)
=\delta_{kl}\delta_{\beta_{k}\gamma_{k}}t_{k\beta_{k}}(j_{\mA}\circ F_{\mathrm{cl}})(Q_{k})\\
&=\delta_{kl}\delta_{\beta_{k}\gamma_{k}}t_{k\beta_{k}}\sum_{i}f_{ki}P_{i}
=\delta_{kl}\delta_{\beta_{k}\gamma_{k}}t_{k\beta_{k}}\sum_{i}f_{ki}\sum_{\alpha_{i}}|e_{i \alpha_{i}}\>\<e_{i \alpha_{i}}|.
\end{split}
\]
By a similar Hilbert--Schmidt adjoint calculation as in the proof of the ($\Leftarrow$) direction, this shows that $\Cd[F]$ is diagonal in these bases. 

Finally, the claim that $[\Cd[F],\rho\otimes1_{\mB}]=0$ follows from these equivalent conditions by simultaneously diagonalizing $\Cd[F]$ and $\rho\otimes1_{\mB}$.
\eprf

\br
The proof of Proposition~\ref{prop:quantumclassical} shows $\sum_{k}f_{ki}=1$ if $F$ is unital and $f_{ki}\ge0$ if $F$ is positive. Hence, if $F$ is positive and unital, then the collection $\{f_{ki}\}$ determines a stochastic matrix. Analogous statements hold for $\rho$ and $\theta$ if $\omega$ and $\omega\circ F$ have analogous properties. 
\er

After Definition~\ref{defn:classicalmodel}, it was claimed that the definition of a classical model was made so as to preserve compositionality. This is stated more precisely in the following. 

\bn
\label{prop:composingclassicalmodels}
Suppose $(\mB\xrightarrow{F}\mA,\mA\xrightarrow{\omega}\C)$ and $(\mC\xrightarrow{G}\mB,\mB\xrightarrow{\omega\circ F}\C)$ have classical models $(\mB_{\mathrm{cl}}\xrightarrow{F_{\mathrm{cl}}}\mA_{\mathrm{cl}},E_{\mA},E_{\mB})$ and $(\mC_{\mathrm{cl}}\xrightarrow{G_{\mathrm{cl}}}\mB_{\mathrm{cl}},E'_{\mB},E_{\mC})$. Then $(\mC\xrightarrow{G\circ F}\mA,\mA\xrightarrow{\omega}\C)$ has $(\mC_{\mathrm{cl}}\xrightarrow{F_{\mathrm{cl}}\circ G_{\mathrm{cl}}}\mA_{\mathrm{cl}},E_{\mA},E_{\mC})$ as a classical model. 
\en

\br
Note that the two conditional expectations $E_{\mB}$ and $E'_{\mB}$ need not be equal in Proposition~\ref{prop:composingclassicalmodels}, but they are almost everywhere equivalent. None of these subtleties arise if all densities have full rank. See~\cite{GPRR21} for details.
\er

\bprf
[Proof of Proposition~\ref{prop:composingclassicalmodels}]
The state-preservation conditions hold by assumption, while 
the factorization follows from
\[
j_{\mA}\circ (F_{\mathrm{cl}}\circ G_{\mathrm{cl}})\circ E_{\mC}
=j_{\mA}\circ F_{\mathrm{cl}}\circ E_{\mB} \circ j_{\mB}\circ G_{\mathrm{cl}}\circ E_{\mC}
=F\circ G.
\qedhere
\]
\eprf

\blem
\label{lem:channelstatesa}
In terms of the notation from Definition~\ref{defn:channelstate}, if $F$ is $\dag$-preserving, then the associated channel density is self-adjoint (equivalently, the channel state is $\dag$-preserving). 
\elem

\bprf
This follows from the fact that $F$ is $\dag$-preserving and $\mu_{\mA}$ is $\dag$-reversing, namely $\mu_{\mA}\circ\dagger=\dagger\circ\mu_{\mA}\circ\gamma$, where $\gamma$ is the swap map. In more detail, 
\[
(\id\otimes F^*)\big(\mu_{\mA}^*(1_{\mA})\big)^{\dag}
=(\id\otimes F^*)\left(\big(\mu_{\mA}^*(1_{\mA})\big)^{\dag}\right)
=(\id\otimes F^*)\Big(\gamma\big(\mu_{\mA}^*(1_{\mA}^{\dag})\big)\Big)
=(\id\otimes F^*)\big(\mu_{\mA}^{*}(1_{\mA})\big).
\]
Equivalently, 
\[
\vcenter{\hbox{
\begin{tikzpicture}[font=\small]
\node[star] (sX) at (-0.5,1.6) {};
\node[star] (sY) at (0.5,1.6) {};
\node[trace] (p) at (0,0) {};
\node[copier] (copier) at (0,0.3) {};
\node[arrow box] (g) at (0.5,0.95) {$F$};
\coordinate (X) at (-0.5,2.0);
\coordinate (Y) at (0.5,2.0);
\draw (p) to (copier);
\draw (copier) to[out=150,in=-90] (sX);
\draw (copier) to[out=15,in=-90] (g);
\draw (g) to (sY);
\draw (sX) to (X);
\draw (sY) to (Y);
\end{tikzpicture}}}
\quad=\quad
\vcenter{\hbox{
\begin{tikzpicture}[font=\small]
\node[trace] (p) at (0,-0.5) {};
\node[copier] (copier) at (0,-0.15) {};
\coordinate (R) at (0.5,0.3) {};
\coordinate (L) at (-0.5,0.3) {};
\coordinate (Ls) at (-0.5,1.1) {};
\coordinate (Rs) at (0.5,1.1) {};
\node[star] (s2) at (-0.5,2.0) {};
\node[star] (s3) at (0.5,2.0) {};
\coordinate (f) at (-0.5,1.4) {};
\node[arrow box] (h) at (0.5,1.4) {$F$};
\coordinate (X) at (0.5,2.3);
\coordinate (Y) at (-0.5,2.3);
\draw (p) to (copier);
\draw (copier) to [out=15,in=-90] (R);
\draw (R) to [out=90,in=-90] (Ls);
\draw (L) to [out=90,in=-90] (Rs);
\draw (Ls) to (f);
\draw (f) to (s2);
\draw (Rs) to (h);
\draw (h) to (s3);
\draw (s3) to (X);
\draw (copier) to[out=165,in=-90] (L);
\draw (s2) to (Y);
\end{tikzpicture}}}
\quad=\quad
\vcenter{\hbox{
\begin{tikzpicture}[font=\small]
\node[trace] (p) at (0,-0.5) {};
\node[copier] (copier) at (0,-0.15) {};
\node[star] (R) at (0.5,0.3) {};
\node[star] (L) at (-0.5,0.3) {};
\coordinate (Ls) at (-0.5,1.4) {};
\coordinate (Rs) at (0.5,1.4) {};
\coordinate (f) at (-0.5,1.7) {};
\node[arrow box] (h) at (0.5,1.7) {$F$};
\coordinate (X) at (0.5,2.25);
\coordinate (Y) at (-0.5,2.25);
\draw (p) to (copier);
\draw (copier) to [out=15,in=-90] (R);
\draw (R) to [out=90,in=-90] (Ls);
\draw (L) to [out=90,in=-90] (Rs);
\draw (Ls) to (f);
\draw (Rs) to (h);
\draw (h) to (X);
\draw (copier) to[out=165,in=-90] (L);
\draw (f) to (Y);
\end{tikzpicture}}}
\quad=\quad
\vcenter{\hbox{
\begin{tikzpicture}[font=\small]
\node[trace] (p) at (0,-0.5) {};
\node[star] (s) at (0,-0.1) {};
\node[copier] (copier) at (0,0.3) {};
\node[arrow box] (g) at (0.5,0.95) {$F$};
\coordinate (X) at (-0.5,1.5);
\coordinate (Y) at (0.5,1.5);
\draw (p) to (s);
\draw (s) to (copier);
\draw (copier) to[out=150,in=-90] (X);
\draw (copier) to[out=15,in=-90] (g);
\draw (g) to (Y);
\end{tikzpicture}}}
\quad=\quad
\vcenter{\hbox{
\begin{tikzpicture}[font=\small]
\node[star] (s) at (0,-0.6) {};
\node[trace] (p) at (0,0) {};
\node[copier] (copier) at (0,0.3) {};
\node[arrow box] (g) at (0.5,0.95) {$F$};
\coordinate (X) at (-0.5,1.5);
\coordinate (Y) at (0.5,1.5);
\draw (p) to (copier);
\draw (copier) to[out=150,in=-90] (X);
\draw (copier) to[out=15,in=-90] (g);
\draw (g) to (Y);
\end{tikzpicture}}}
\quad,
\]
where the first identity follows from the properties of the trace, and where $\,\vcenter{\hbox{%
\begin{tikzpicture}[font=\footnotesize]
\node[star] (s) at (0,0.3) {};
\draw (s) to (0,0.6);
\draw (s) to (0,0);
\end{tikzpicture}}}\,$
denotes the involution (cf.\ \cite{PaBayes}).
\eprf

\blem
\label{lem:jordanbloomdensities}
Given linear maps $F:\mB\to\mA$ and $\omega:\mA\to\C$, 
the densities associated with $\omega\circ\bloom_{F}$ and $\omega\circ\Lbloom{F}$ are $\Cd[F](\rho\otimes1_{\mB})$ and $(\rho\otimes1_{\mB})\Cd[F]$, respectively, where $\rho:=\Cd[\omega]$.  
\elem

\bprf
We prove one of these claims as the other is completely analogous. Indeed, by using the cyclicity property of the trace and the definition of the Hilbert--Schmidt adjoint, 
\[
\begin{split}
\tr\Big(\Cd[F](\rho\otimes1_{\mB})(A\otimes B)\Big)
&=\tr\Big(\big((\id_{\mA}\otimes F^*)\mu_{\mA}^*(1_{\mA})\big)^{\dag}(\rho A\otimes B)\Big)\\
&=\tr\Big(\mu_{\mA}^*(1_{\mA})^{\dag} \big(\rho A\otimes F(B)\big)\Big)\\
&=\tr\Big(1_{\mA}^{\dag}\mu_{\mA}\big(\rho A\otimes F(B)\big)\Big)\\
&=\tr\big(\rho A F(B)\big)\\
&=(\omega\circ\bloom_{F})(A\otimes B).
\end{split}
\]
Since $A\in\mA$ and $B\in\mB$ are arbitrary and since the trace is non-degenerate, the density associated with $\omega\circ\bloom_{F}$ is $\Cd[F](\rho\otimes1_{\mB})$. 
\eprf

\blem
\label{lem:jordansa}
Given unital $\dag$-preserving maps $F:\mB\to\mA$ and $\omega:\mA\to\C$, one has
\[
\omega\circ\bloom_{F}\circ\dag=\dag\circ\omega\circ\Lbloom{F}
\quad
\text{ and }
\quad
\omega\circ\Lbloom{F}\circ\dag=\dag\circ\omega\circ\bloom_{F}.
\]
In particular, the functional $F\star\omega:=\frac{1}{2}(\omega\circ\bloom_{F}+\omega\circ\Lbloom{F})$ is unital and $\dag$-preserving and has a density given by the Jordan product of the channel density together with the density $\rho:=\Cd[\omega]$, i.e., 
\[
\Cd[F\star\omega]=\frac{1}{2}\Big(\Cd[F](\rho\otimes1_{\mB})+(\rho\otimes1_{\mB})\Cd[F]\Big).
\]
\elem

\bprf
The first two identities follow from Lemma~\ref{lem:channelstatesa} and Lemma~\ref{lem:jordanbloomdensities}. Namely, 
\[
\Cd[\omega\circ\bloom_{F}]^{\dag}
=\big(\Cd[F](\rho\otimes1_{\mB})\big)^{\dag}
=(\rho\otimes1_{\mB})^{\dag}\Cd[F]^{\dag}
=(\rho\otimes1_{\mB})\Cd[F]
=\Cd[\omega\circ\Lbloom{F}].
\]
The $\dag$-preserving property follows from this.
The unitality follows from the fact that the composite of unital maps is unital. The formula for the density in terms of the Jordan product follows from the first two identities and Lemma~\ref{lem:jordanbloomdensities}.
\eprf

\blem
\label{lem:marginals}
Let $F:\mB\to\mA$ and $\omega:\mA\to\C$ be linear maps, with $F$ unital. Then 
\[
\omega\circ\bloom_{F}\circ i_{\mA}=\omega=\omega\circ\Lbloom{F}\circ i_{\mA}
\qquad\text{ and }\qquad
\omega\circ\bloom_{F}\circ i_{\mB}=\omega\circ F=\omega\circ\Lbloom{F}\circ i_{\mB}.
\]
\elem

\bprf
The proofs are straightforward calculations. For example, 
\[
(\omega\circ\bloom_{F}\circ i_{\mA})(A)
=(\omega\circ\mu_{\mA}\circ(\id_{\mA}\otimes F)\circ i_{\mA})(A)
=\omega\big(AF(1_{\mB})\big)
=\omega(A1_{\mA})
=\omega(A)
\]
and
\[
(\omega\circ\Lbloom{F}\circ i_{\mA})(A)
=\big(\omega\circ\mu_{\mA}\circ(F\otimes\id_{\mA})\circ\gamma\circ i_{\mA}\big)(A)
=\omega\Big(\mu_{\mA}\big(F(1_{\mB})A\big)\Big)
=\omega\big(F(1_{\mB})A\big)
=\omega(A)
\]
for all $A\in\mA$. A similar calculation holds for the second set of claims. The calculations are easily visualized using string diagrams. For example, the two we just showed are given by
\[
\vcenter{\hbox{
\begin{tikzpicture}[font=\small]
\node[state] (p) at (0,-0.1) {$\omega$};
\node[copier] (copier) at (0,0.3) {};
\node[arrow box] (g) at (0.5,0.95) {$F$};
\coordinate (X) at (-0.5,1.8);
\node[discarder] (Y) at (0.5,1.5) {};
\draw (p) to (copier);
\draw (copier) to[out=150,in=-90] (X);
\draw (copier) to[out=15,in=-90] (g);
\draw (g) to (Y);
\end{tikzpicture}}}
\quad
=
\quad
\vcenter{\hbox{%
\begin{tikzpicture}[font=\small]
\node[state] (p) at (0,0) {$\omega$};
\node (X) at (0,0.7) {};
\draw (p) to (X);
\end{tikzpicture}}}
\quad
=
\quad
\vcenter{\hbox{
\begin{tikzpicture}[font=\small]
\node[state] (p) at (0,-0.5) {$\omega$};
\node[copier] (copier) at (0,-0.15) {};
\coordinate (R) at (0.5,0.3) {};
\coordinate (Ls) at (-0.5,1.6) {};
\coordinate (Rs) at (0.5,1.6) {};
\coordinate (s2) at (-0.5,1.7) {};
\coordinate (s3) at (0.5,1.7) {};
\coordinate (g) at (0.5,0.7) {};
\node[arrow box] (h) at (-0.5,0.5) {$F$};
\node[discarder] (X) at (0.5,1.7) {};
\coordinate (Y) at (-0.5,2.1);
\draw (p) to (copier);
\draw (copier) to [out=15,in=-90] (g);
\draw (g) to [out=90,in=-90] (Ls);
\draw (h) to [out=90,in=-90] (Rs);
\draw (Ls) to (s2);
\draw (Rs) to (s3);
\draw (s3) to (X);
\draw (copier) to[out=165,in=-90] (h);
\draw (s2) to (Y);
\end{tikzpicture}}}
.
\]
Note that only unitality of $F$ was used here.
\eprf

\bprf
[Proof of Theorem~\ref{thm:statesovertime}]
{\color{white}{you found me!}}

\begin{enumerate}[(a)]
\itemsep0em
\item
This follows from Lemma~\ref{lem:jordansa}. 
\item
This follows from the linearity of $\bloom_{F}$ and $\Lbloom{F}$ in the argument $F$ as well as linearity of $\omega$. Indeed, writing $F\star\omega$ out more explicitly as
\[
F\star\omega=\frac{1}{2}\Big(\omega\circ\mu_{\mA}\circ(\id\otimes F)+\omega\circ\mu_{\mA}\circ(F\otimes\id)\circ\gamma\Big)
\]
shows that in fact each term on the right-hand-side depends linearly on both $F$ and $\omega$.
\item
In what follows, we will prove $\omega\circ\Lbloom{F}=\omega\circ\bloom_{F}$. We will use string diagrams for an elegant proof. It will use the fact that state-preserving conditional expectations are Bayesian inverses~\cite{GPRR21}, and it will also crucially use the fact that $\mu_{\mA_{\mathrm{cl}}}\circ\gamma=\mu_{\mA_{\mathrm{cl}}}$, which only holds for commutative $C^*$-algebras. We implement the notation of Definition~\ref{defn:classicalmodel}. The calculation 
\[
\vcenter{\hbox{
\begin{tikzpicture}[font=\small]
\node[state] (p) at (0,-0.5) {$\omega$};
\node[copier] (copier) at (0,-0.15) {};
\coordinate (R) at (0.5,0.3) {};
\coordinate (Ls) at (-0.5,1.6) {};
\coordinate (Rs) at (0.5,1.6) {};
\coordinate (s2) at (-0.5,1.7) {};
\coordinate (s3) at (0.5,1.7) {};
\coordinate (g) at (0.5,0.7) {};
\node[arrow box] (h) at (-0.5,0.5) {$F$};
%
\draw (p) to (copier);
\draw (copier) to [out=15,in=-90] (g);
\draw (g) to [out=90,in=-90] (Ls);
\draw (h) to [out=90,in=-90] (Rs);
\draw (Ls) to (s2);
\draw (Rs) to (s3);
\draw (copier) to[out=165,in=-90] (h);
\end{tikzpicture}}}
\;
=
\;
\vcenter{\hbox{
\begin{tikzpicture}[font=\small]
\node[state] (p) at (0,-0.5) {$\omega$};
\node[copier] (copier) at (0,-0.15) {};
\coordinate (R) at (0.5,0.3) {};
\coordinate (Ls) at (-0.5,3.6) {};
\coordinate (Rs) at (0.5,3.6) {};
\coordinate (s2) at (-0.5,3.7) {};
\coordinate (s3) at (0.5,3.7) {};
\coordinate (g1) at (0.5,0.5) {};
\coordinate (g2) at (0.5,2.7) {};
\node[arrow box] (h1) at (-0.5,0.5) {$j_{\mA}$};
\node[arrow box] (h2) at (-0.5,1.5) {$F_{\mathrm{cl}}$};
\node[arrow box] (h3) at (-0.5,2.5) {$E_{\mB}$};
%
\draw (p) to (copier);
\draw (copier) to [out=15,in=-90] (g1);
\draw (g1) to (g2);
\draw (g2) to [out=90,in=-90] (Ls);
\draw (h3) to [out=90,in=-90] (Rs);
\draw (Ls) to (s2);
\draw (Rs) to (s3);
\draw (copier) to[out=165,in=-90] (h1);
\draw (h1) to (h2);
\draw (h2) to (h3);
\end{tikzpicture}}}
\;
=
\;
\vcenter{\hbox{
\begin{tikzpicture}[font=\small]
\node[state] (p) at (0,-0.5) {$\omega_{\restriction}$};
\node[copier] (copier) at (0,-0.15) {};
\coordinate (R) at (0.5,0.3) {};
\coordinate (Ls) at (-0.5,2.6) {};
\coordinate (Rs) at (0.5,2.6) {};
\coordinate (s2) at (-0.5,2.7) {};
\coordinate (s3) at (0.5,2.7) {};
\node[arrow box] (g1) at (0.5,0.5) {$E_{\mA}$};
\coordinate (g2) at (0.5,1.7) {};
\node[arrow box] (h2) at (-0.5,0.5) {$F_{\mathrm{cl}}$};
\node[arrow box] (h3) at (-0.5,1.5) {$E_{\mB}$};
%
\draw (p) to (copier);
\draw (copier) to [out=15,in=-90] (g1);
\draw (g1) to (g2);
\draw (g2) to [out=90,in=-90] (Ls);
\draw (h3) to [out=90,in=-90] (Rs);
\draw (Ls) to (s2);
\draw (Rs) to (s3);
\draw (copier) to[out=165,in=-90] (h2);
\draw (h2) to (h3);
\end{tikzpicture}}}
\;
=
\;
\vcenter{\hbox{
\begin{tikzpicture}[font=\small]
\node[state] (p) at (0,-0.5) {$\omega_{\restriction}$};
\node[copier] (copier) at (0,-0.15) {};
\coordinate (R) at (0.5,0.3) {};
\coordinate (L) at (-0.5,0.3) {};
\coordinate (Ls) at (-0.5,1.4) {};
\coordinate (Rs) at (0.5,1.4) {};
\coordinate (s2) at (-0.5,2.7) {};
\node[arrow box] (s3) at (0.5,2.7) {$E_{\mB}$};
\node[arrow box] (f) at (-0.5,1.7) {$E_{\mA}$};
\node[arrow box] (h) at (0.5,1.7) {$F_{\mathrm{cl}}$};
\coordinate (X) at (0.5,3.4);
\coordinate (Y) at (-0.5,3.4);
\draw (p) to (copier);
\draw (copier) to [out=15,in=-90] (R);
\draw (R) to [out=90,in=-90] (Ls);
\draw (L) to [out=90,in=-90] (Rs);
\draw (Ls) to (f);
\draw (f) to (s2);
\draw (Rs) to (h);
\draw (h) to (s3);
\draw (s3) to (X);
\draw (copier) to[out=165,in=-90] (L);
\draw (s2) to (Y);
\end{tikzpicture}}}
\;
=
\;
\vcenter{\hbox{%
\begin{tikzpicture}[font=\small]
\node[copier] (copier) at (0,0.3) {};
\node[arrow box] (h) at (-0.5,0.95) {$E_{\mA}$};
\node[arrow box] (f) at (0.5,0.95) {$F_{\mathrm{cl}}$};
\node[arrow box] (f2) at (0.5,1.95) {$E_{\mB}$};
\node[state] (p) at (0,-0) {$\omega_{\restriction}$};
\coordinate (X) at (-0.5,2.6);
\coordinate (Y) at (0.5,2.6);
\draw (p) to (copier);
\draw (copier) to[out=165,in=-90] (h);
\draw (h) to (X);
\draw (copier) to[out=15,in=-90] (f);
\draw (f) to (f2);
\draw (f2) to (Y);
\end{tikzpicture}}}
\;
=
\;
\vcenter{\hbox{
\begin{tikzpicture}[font=\small,xscale=-1]
\node[state] (p) at (0,-0.5) {$\omega$};
\node[copier] (copier) at (0,-0.15) {};
\coordinate (R) at (0.5,0.3) {};
\coordinate (s2) at (-0.5,3.1) {};
\coordinate (s3) at (0.5,3.1) {};
\coordinate (g1) at (0.5,0.5) {};
\coordinate (g2) at (0.5,2.7) {};
\node[arrow box] (h1) at (-0.5,0.5) {$j_{\mA}$};
\node[arrow box] (h2) at (-0.5,1.5) {$F_{\mathrm{cl}}$};
\node[arrow box] (h3) at (-0.5,2.5) {$E_{\mB}$};
%
\draw (p) to (copier);
\draw (copier) to [out=15,in=-90] (g1);
\draw (g1) to (g2);
\draw (g2) to [out=90,in=-90] (s3);
\draw (h3) to [out=90,in=-90] (s2);
\draw (copier) to[out=165,in=-90] (h1);
\draw (h1) to (h2);
\draw (h2) to (h3);
\end{tikzpicture}}}
\;
=
\;
\vcenter{\hbox{
\begin{tikzpicture}[font=\small]
\node[state] (p) at (0,0) {$\omega$};
\node[copier] (copier) at (0,0.3) {};
\node[arrow box] (g) at (0.5,0.95) {$F$};
\coordinate (X) at (-0.5,1.5);
\coordinate (Y) at (0.5,1.5);
\draw (p) to (copier);
\draw (copier) to[out=150,in=-90] (X);
\draw (copier) to[out=15,in=-90] (g);
\draw (g) to (Y);
\end{tikzpicture}}}
\]
implies the claim. 
\item
This follows from Lemma~\ref{lem:marginals}. 
\item
The proof of associativity will be achieved by showing commutativity of the diagram in Remark~\ref{rmk:associativity}. Following along the left-hand-side of that diagram results in 
\begingroup
\allowdisplaybreaks
\begin{align*}
\left(
\vcenter{\hbox{%
\begin{tikzpicture}[font=\small]
\node[arrow box] (c) at (0,0) {$G$};
\coordinate (d) at (0,0.7) {};
\draw  (c) to (d);
\draw (c) to (0,-0.7);
\end{tikzpicture}}}
\;,\;
\vcenter{\hbox{%
\begin{tikzpicture}[font=\small]
\node[arrow box] (c) at (0,0) {$F$};
\coordinate (d) at (0,0.7) {};
\draw  (c) to (d);
\draw (c) to (0,-0.7);
\end{tikzpicture}}}
\;,\;
\vcenter{\hbox{%
\begin{tikzpicture}[font=\small]
\node[state] (p) at (0,0) {$\omega$};
\node (X) at (0,0.7) {};
\draw (p) to (X);
\end{tikzpicture}}}
\right)
&\xmapsto{(\shriek_{\mA}\otimes\;\cdot\;)\times\star}
\left(
\vcenter{\hbox{%
\begin{tikzpicture}[font=\small]
\node[discarder] (A) at (-0.8,0.2) {};
\coordinate (X) at (-0.8,-0.7) {};
\node[arrow box] (c) at (0,0) {$G$};
\coordinate (d) at (0,0.7) {};
\draw  (c) to (d);
\draw (c) to (0,-0.7);
\draw (X) to (A);
\end{tikzpicture}}}
\quad,\quad
\frac{1}{2}
\left(
\vcenter{\hbox{
\begin{tikzpicture}[font=\small]
\node[state] (p) at (0,0) {$\omega$};
\node[copier] (copier) at (0,0.3) {};
\node[arrow box] (g) at (0.5,0.95) {$F$};
\coordinate (X) at (-0.5,1.5);
\coordinate (Y) at (0.5,1.5);
\draw (p) to (copier);
\draw (copier) to[out=150,in=-90] (X);
\draw (copier) to[out=15,in=-90] (g);
\draw (g) to (Y);
\end{tikzpicture}}}
\;
+
\;
\vcenter{\hbox{
\begin{tikzpicture}[font=\small]
\node[state] (p) at (0,-0.5) {$\omega$};
\node[copier] (copier) at (0,-0.15) {};
\coordinate (R) at (0.5,0.3) {};
\coordinate (Ls) at (-0.5,1.6) {};
\coordinate (Rs) at (0.5,1.6) {};
\coordinate (s2) at (-0.5,1.7) {};
\coordinate (s3) at (0.5,1.7) {};
\coordinate (g) at (0.5,0.7) {};
\node[arrow box] (h) at (-0.5,0.5) {$F$};
%
\draw (p) to (copier);
\draw (copier) to [out=15,in=-90] (g);
\draw (g) to [out=90,in=-90] (Ls);
\draw (h) to [out=90,in=-90] (Rs);
\draw (Ls) to (s2);
\draw (Rs) to (s3);
\draw (copier) to[out=165,in=-90] (h);
\end{tikzpicture}}}
\right)
\right)
\\
&\xmapsto{\star}
\frac{1}{4}\left(
\vcenter{\hbox{
\begin{tikzpicture}[font=\small,xscale=-1]
\node[state] (q) at (0,-0.1) {$\omega$};
\node[copier] (c) at (0,0.3) {};
\node[copier] (c2) at (-0.5,1.55) {};
\node[copier] (c3) at (0.5,1.55) {};
\node[arrow box] (g) at (-0.5,0.95) {$F$};
\node[arrow box] (f) at (-1.1,2.25) {$G$};
\coordinate (X) at (1.0,2.85);
\coordinate (Y1) at (0.1,2.85);
\coordinate (Y2) at (-1.1,2.85);
\node[discarder] (d) at (-0.5,2.3) {};
\draw (q) to (c);
\draw (c) to [out=15,in=-90] (c3);
\draw (c3) to [out=15,in=-90] (X);
\draw (c) to [out=165,in=-90] (g);
\draw (c2) to [out=165,in=-90] (f);
\draw (f) to (Y2);
\draw (c2) to[out=15,in=-90] (Y1);
\draw (g) to (c2);
\draw (c3) to [out=165,in=-90] (d);
\end{tikzpicture}}}
+
\vcenter{\hbox{
\begin{tikzpicture}[font=\small,xscale=-1]
\node[state] (q) at (0,0) {$\omega$};
\node[copier] (c) at (0,0.3) {};
\node[copier] (c3) at (0.5,1.55) {};
\node[discarder] (d) at (1.0,2.3) {};
\node[copier] (c2) at (-0.5,1.55) {};
\node[arrow box] (g) at (-0.5,0.95) {$F$};
\node[arrow box] (f) at (0.1,2.3) {$G$};
\coordinate (M) at (-0.5,2.3);
\coordinate (Y1) at (0.1,2.7);
\coordinate (Y2) at (-1.1,2.7);
\coordinate (C) at (0.8,3.9);
\coordinate (B) at (0.1,3.9);
\coordinate (A) at (-1.1,3.9);
\draw (q) to (c);
\draw (c) to[out=15,in=-90] (c3);
\draw (c3) to[out=165,in=-90] (M);
\draw (c3) to[out=15,in=-90] (d);
\draw (M) to [out=90,in=-90] (C);
\draw (c) to[out=165,in=-90] (g);
\draw (c2) to[out=15,in=-90] (f);
\draw (f) to (Y1);
\draw (c2) to[out=165,in=-90] (Y2);
\draw (g) to (c2);
\draw (Y1) to [out=90,in=-90] (A);
\draw (Y2) to [out=90,in=-90] (B);
\end{tikzpicture}}}
+
\vcenter{\hbox{
\begin{tikzpicture}[font=\small]
\node[state] (p) at (0,-0.5) {$\omega$};
\node[copier] (copier) at (0,-0.15) {};
\coordinate (R) at (0.5,0.3) {};
\coordinate (Ls) at (-0.5,1.6) {};
\coordinate (Rs) at (0.5,1.6) {};
\node[copier] (s2) at (-0.5,1.9) {};
\coordinate (s3) at (0.5,1.7) {};
\coordinate (g) at (0.5,0.7) {};
\node[arrow box] (h) at (-0.5,0.5) {$F$};
\node[copier] (X) at (0.5,1.9) {};
\node[discarder] (d) at (0.4,2.7) {};
\node[arrow box] (G) at (1.1,2.7) {$G$};
\coordinate (B) at (-0.2,3.3);
\coordinate (Y) at (-1.0,3.3);
\coordinate (C) at (1.1,3.3);
\draw (p) to (copier);
\draw (copier) to [out=15,in=-90] (g);
\draw (g) to [out=90,in=-90] (Ls);
\draw (h) to [out=90,in=-90] (Rs);
\draw (Ls) to (s2);
\draw (Rs) to (s3);
\draw (s3) to (X);
\draw (copier) to[out=165,in=-90] (h);
\draw (s2) to[out=165,in=-90] (Y);
\draw (s2) to[out=15,in=-90] (d);
\draw (X) to [out=15,in=-90] (G);
\draw (X) to[out=165,in=-90] (B);
\draw (G) to (C); 
\end{tikzpicture}}}
+
\vcenter{\hbox{
\begin{tikzpicture}[font=\small]
\node[state] (q) at (-0.5,-2) {$\omega$};
\node[copier] (c) at (-0.5,-1.6) {};
\node[arrow box] (p) at (-1,-0.9) {$F$};
\coordinate (r) at (0,-0.7) {};
\node[copier] (copier) at (0,0.25) {};
\node[copier] (cL) at (-1.0,0.25) {};
\coordinate (M) at (0,0.9) {};
\coordinate (R) at (0.5,0.7) {};
\coordinate (Ls) at (-0.5,2.4) {};
\coordinate (Rs) at (0.5,2.4) {};
\coordinate (s2) at (-0.5,2.5) {};
\coordinate (s3) at (0.5,2.5) {};
\coordinate (g) at (0.5,1.1) {};
\node[arrow box] (h) at (-0.6,1.0) {$G$};
\node[discarder] (d) at (-1.5,1.1) {};
\coordinate (X) at (0.5,2.5);
\coordinate (Y) at (-0.5,2.5);
\coordinate (A) at (-1.3,2.5);
\draw (q) to (c);
\draw (c) to[out=165,in=-90] (p);
\draw (cL) to[out=165,in=-90] (d);
\draw (cL) to[out=15,in=-90] (M);
\draw (M) to[out=90,in=-90] (A);
\draw (c) to[out=15,in=-90] (r);
\draw (p) to[out=90,in=-90] (copier);
\draw (r) to[out=90,in=-90] (cL);
\draw (copier) to [out=15,in=-90] (g);
\draw (g) to [out=90,in=-90] (Ls);
\draw (h) to [out=90,in=-90] (Rs);
\draw (Ls) to (s2);
\draw (Rs) to (s3);
\draw (s3) to (X);
\draw (copier) to[out=165,in=-90] (h);
\draw (s2) to (Y);
\end{tikzpicture}}}
\right)
\\
&=
\frac{1}{4}\left(
\vcenter{\hbox{
\begin{tikzpicture}[font=\small,xscale=-1]
\node[state] (q) at (0,-0.1) {$\omega$};
\node[copier] (c) at (0,0.3) {};
\node[copier] (c2) at (-0.5,1.55) {};
\node[arrow box] (g) at (-0.5,0.95) {$F$};
\node[arrow box] (f) at (-1,2.25) {$G$};
\coordinate (X) at (0.7,2.85);
\coordinate (Y1) at (0,2.85);
\coordinate (Y2) at (-1,2.85);
\draw (q) to (c);
\draw (c) to[out=15,in=-90] (X);
\draw (c) to[out=165,in=-90] (g);
\draw (c2) to[out=165,in=-90] (f);
\draw (f) to (Y2);
\draw (c2) to[out=15,in=-90] (Y1);
\draw (g) to (c2);
\end{tikzpicture}}}
+
\vcenter{\hbox{
\begin{tikzpicture}[font=\small,xscale=-1]
\node[state] (q) at (0,0) {$\omega$};
\node[copier] (c) at (0,0.3) {};
\node[copier] (c2) at (-0.5,1.55) {};
\node[arrow box] (g) at (-0.5,0.95) {$F$};
\node[arrow box] (f) at (0,2.25) {$G$};
\coordinate (X) at (0.7,2.7);
\coordinate (Y1) at (0,2.7);
\coordinate (Y2) at (-1,2.7);
\coordinate (C) at (0.7,3.9);
\coordinate (B) at (0,3.9);
\coordinate (A) at (-1,3.9);
\draw (q) to (c);
\draw (c) to[out=15,in=-90] (X);
\draw (c) to[out=165,in=-90] (g);
\draw (c2) to[out=15,in=-90] (f);
\draw (f) to (Y1);
\draw (c2) to[out=165,in=-90] (Y2);
\draw (g) to (c2);
\draw (X) to [out=90,in=-90] (C);
\draw (Y1) to [out=90,in=-90] (A);
\draw (Y2) to [out=90,in=-90] (B);
\end{tikzpicture}}}
+
\vcenter{\hbox{
\begin{tikzpicture}[font=\small]
\node[state] (p) at (0,-0.5) {$\omega$};
\node[copier] (copier) at (0,-0.15) {};
\coordinate (R) at (0.5,0.3) {};
\coordinate (Ls) at (-0.5,1.6) {};
\coordinate (Rs) at (0.5,1.6) {};
\coordinate (s2) at (-0.5,1.7) {};
\coordinate (s3) at (0.5,1.7) {};
\coordinate (g) at (0.5,0.7) {};
\node[arrow box] (h) at (-0.5,0.5) {$F$};
\node[copier] (X) at (0.5,1.9) {};
\node[arrow box] (G) at (1.0,2.7) {$G$};
\coordinate (B) at (0,3.5);
\coordinate (Y) at (-0.5,3.5);
\coordinate (C) at (1.0,3.5);
\draw (p) to (copier);
\draw (copier) to [out=15,in=-90] (g);
\draw (g) to [out=90,in=-90] (Ls);
\draw (h) to [out=90,in=-90] (Rs);
\draw (Ls) to (s2);
\draw (Rs) to (s3);
\draw (s3) to (X);
\draw (copier) to[out=165,in=-90] (h);
\draw (s2) to (Y);
\draw (X) to [out=15,in=-90] (G);
\draw (X) to[out=165,in=-90] (B);
\draw (G) to (C); 
\end{tikzpicture}}}
+
\vcenter{\hbox{
\begin{tikzpicture}[font=\small]
\node[state] (p) at (0,-0.5) {$\omega$};
\node[copier] (copier) at (0,-0.15) {};
\coordinate (R) at (0.5,0.3) {};
\coordinate (Ls) at (-0.5,1.6) {};
\coordinate (Rs) at (0.5,1.6) {};
\coordinate (s2) at (-0.5,1.7) {};
\coordinate (s3) at (0.5,1.7) {};
\coordinate (g) at (0.5,0.7) {};
\node[arrow box] (h) at (-0.5,0.5) {$F$};
\node[copier] (X) at (0.5,1.9) {};
\node[arrow box] (G) at (0,2.7) {$G$};
\coordinate (Y) at (-0.5,3.9);
\coordinate (C) at (1.0,2.9);
\coordinate (B2) at (0,3.9);
\coordinate (C2) at (1.0,3.9);
\draw (p) to (copier);
\draw (copier) to [out=15,in=-90] (g);
\draw (g) to [out=90,in=-90] (Ls);
\draw (h) to [out=90,in=-90] (Rs);
\draw (Ls) to (s2);
\draw (Rs) to (s3);
\draw (s3) to (X);
\draw (copier) to[out=165,in=-90] (h);
\draw (s2) to (Y);
\draw (X) to [out=165,in=-90] (G);
\draw (X) to[out=15,in=-90] (C);
\draw (G) to [out=90,in=-90] (C2);
\draw (C) to [out=90,in=-90] (B2);
\end{tikzpicture}}}
\right)
.
\end{align*}
\endgroup
Meanwhile, following along the top and right side of the diagram in Remark~\ref{rmk:associativity} gives
\begingroup
\allowdisplaybreaks
\begin{align*}
\left(
\vcenter{\hbox{%
\begin{tikzpicture}[font=\small]
\node[arrow box] (c) at (0,0) {$G$};
\coordinate (d) at (0,0.7) {};
\draw  (c) to (d);
\draw (c) to (0,-0.7);
\end{tikzpicture}}}
\;,\;
\vcenter{\hbox{%
\begin{tikzpicture}[font=\small]
\node[arrow box] (c) at (0,0) {$F$};
\coordinate (d) at (0,0.7) {};
\draw  (c) to (d);
\draw (c) to (0,-0.7);
\end{tikzpicture}}}
\;,\;
\vcenter{\hbox{%
\begin{tikzpicture}[font=\small]
\node[state] (p) at (0,0) {$\omega$};
\node (X) at (0,0.7) {};
\draw (p) to (X);
\end{tikzpicture}}}
\right)
&\xmapsto{(\shriek_{\mA}\otimes\;\cdot\;)\times\Cs_{\mA,\mB}\times\id}
\left(
\vcenter{\hbox{%
\begin{tikzpicture}[font=\small]
\node[discarder] (A) at (-0.8,0.2) {};
\coordinate (X) at (-0.8,-0.7) {};
\node[arrow box] (c) at (0,0) {$G$};
\coordinate (d) at (0,0.7) {};
\draw  (c) to (d);
\draw (c) to (0,-0.7);
\draw (X) to (A);
\end{tikzpicture}}}
\quad,\quad
\vcenter{\hbox{
\begin{tikzpicture}[font=\small]
\node[trace] (p) at (0,0) {};
\node[copier] (copier) at (0,0.3) {};
\node[arrow box] (g) at (0.5,0.95) {$F$};
\coordinate (X) at (-0.5,1.5);
\coordinate (Y) at (0.5,1.5);
\draw (p) to (copier);
\draw (copier) to[out=150,in=-90] (X);
\draw (copier) to[out=15,in=-90] (g);
\draw (g) to (Y);
\end{tikzpicture}}}
\quad,\quad
\vcenter{\hbox{%
\begin{tikzpicture}[font=\small]
\node[state] (p) at (0,0) {$\omega$};
\node (X) at (0,0.7) {};
\draw (p) to (X);
\end{tikzpicture}}}
\right)
\\
&\xmapsto{\star\times\id}
\left(
\frac{1}{2}\left(
\vcenter{\hbox{
\begin{tikzpicture}[font=\small,xscale=-1]
\node[trace] (q) at (0,0) {};
\node[copier] (c) at (0,0.3) {};
\node[copier] (c2) at (-0.5,1.55) {};
\node[arrow box] (g) at (-0.5,0.95) {$F$};
\node[arrow box] (f) at (-1,2.25) {$G$};
\coordinate (X) at (0.7,2.85);
\coordinate (Y1) at (0,2.85);
\coordinate (Y2) at (-1,2.85);
\draw (q) to (c);
\draw (c) to[out=15,in=-90] (X);
\draw (c) to[out=165,in=-90] (g);
\draw (c2) to[out=165,in=-90] (f);
\draw (f) to (Y2);
\draw (c2) to[out=15,in=-90] (Y1);
\draw (g) to (c2);
\end{tikzpicture}}}
+
\vcenter{\hbox{
\begin{tikzpicture}[font=\small]
\node[trace] (q) at (-0.5,-1.9) {};
\node[copier] (c) at (-0.5,-1.6) {};
\node[arrow box] (p) at (0,-0.8) {$F$};
\node[copier] (copier) at (0,-0.15) {};
\node[copier] (cL) at (-1.0,-0.15) {};
\coordinate (M) at (0,0.5) {};
\coordinate (R) at (0.5,0.3) {};
\coordinate (g) at (0.5,0.7) {};
\node[arrow box] (h) at (-0.6,0.6) {$G$};
\node[discarder] (d) at (-1.5,0.7) {};
\coordinate (X) at (0.5,1.9);
\coordinate (Y) at (-0.5,1.9);
\coordinate (A) at (-1.3,1.9);
\draw (q) to (c);
\draw (c) to[out=165,in=-90] (cL);
\draw (cL) to[out=165,in=-90] (d);
\draw (cL) to[out=15,in=-90] (M);
\draw (M) to[out=90,in=-90] (A);
\draw (c) to[out=15,in=-90] (p);
\draw (p) to (copier);
\draw (copier) to [out=15,in=-90] (g);
\draw (g) to [out=90,in=-90] (Y);
\draw (h) to [out=90,in=-90] (X);
\draw (copier) to[out=165,in=-90] (h);
\end{tikzpicture}}}
\right)
\quad,\quad
\vcenter{\hbox{%
\begin{tikzpicture}[font=\small]
\node[state] (p) at (0,0) {$\omega$};
\node (X) at (0,0.7) {};
\draw (p) to (X);
\end{tikzpicture}}}
\right)
\\
&=
\left(
\frac{1}{2}
\left(
\vcenter{\hbox{
\begin{tikzpicture}[font=\small,xscale=-1]
\node[trace] (q) at (0,0) {};
\node[copier] (c) at (0,0.3) {};
\node[copier] (c2) at (-0.5,1.55) {};
\node[arrow box] (g) at (-0.5,0.95) {$F$};
\node[arrow box] (f) at (-1,2.25) {$G$};
\coordinate (X) at (0.7,2.85);
\coordinate (Y1) at (0,2.85);
\coordinate (Y2) at (-1,2.85);
\draw (q) to (c);
\draw (c) to[out=15,in=-90] (X);
\draw (c) to[out=165,in=-90] (g);
\draw (c2) to[out=165,in=-90] (f);
\draw (f) to (Y2);
\draw (c2) to[out=15,in=-90] (Y1);
\draw (g) to (c2);
\end{tikzpicture}}}
+
\vcenter{\hbox{
\begin{tikzpicture}[font=\small]
\node[trace] (q) at (-0.5,-1.9) {};
\node[copier] (c) at (-0.5,-1.6) {};
\node[arrow box] (p) at (0,-0.8) {$F$};
\node[copier] (copier) at (0,-0.15) {};
\coordinate (R) at (0.5,0.3) {};
\coordinate (Ls) at (-0.5,1.6) {};
\coordinate (Rs) at (0.5,1.6) {};
\coordinate (s2) at (-0.5,1.7) {};
\coordinate (s3) at (0.5,1.7) {};
\coordinate (g) at (0.5,0.7) {};
\node[arrow box] (h) at (-0.5,0.5) {$G$};
\coordinate (X) at (0.5,1.7);
\coordinate (Y) at (-0.5,1.7);
\coordinate (A) at (-1.3,1.7);
\draw (q) to (c);
\draw (c) to[out=165,in=-90] (A);
\draw (c) to[out=15,in=-90] (p);
\draw (p) to (copier);
\draw (copier) to [out=15,in=-90] (g);
\draw (g) to [out=90,in=-90] (Ls);
\draw (h) to [out=90,in=-90] (Rs);
\draw (Ls) to (s2);
\draw (Rs) to (s3);
\draw (s3) to (X);
\draw (copier) to[out=165,in=-90] (h);
\draw (s2) to (Y);
\end{tikzpicture}}}
\right)
\quad,\quad
\vcenter{\hbox{%
\begin{tikzpicture}[font=\small]
\node[state] (p) at (0,0) {$\omega$};
\node (X) at (0,0.7) {};
\draw (p) to (X);
\end{tikzpicture}}}
\right)
\\
&\xmapsto{\Cs_{\mA,\mB\otimes\mC}^{-1}\times\id}
\left(
\frac{1}{2}
\left(
\vcenter{\hbox{
\begin{tikzpicture}[font=\small,xscale=-1]
\node[arrow box] (p) at (0,-0.3) {$F$};
\node[copier] (copier) at (0,0.3) {};
\node[arrow box] (g) at (-0.5,0.95) {$G$};
\coordinate (X) at (0.5,1.5);
\coordinate (Y) at (-0.5,1.5);
\draw (0,-0.9) to (p);
\draw (p) to (copier);
\draw (copier) to[out=15,in=-90] (X);
\draw (copier) to[out=165,in=-90] (g);
\draw (g) to (Y);
\end{tikzpicture}}}
+
\vcenter{\hbox{
\begin{tikzpicture}[font=\small]
\node[arrow box] (p) at (0,-0.8) {$F$};
\node[copier] (copier) at (0,-0.15) {};
\coordinate (R) at (0.5,0.3) {};
\coordinate (Ls) at (-0.5,1.6) {};
\coordinate (Rs) at (0.5,1.6) {};
\coordinate (s2) at (-0.5,1.7) {};
\coordinate (s3) at (0.5,1.7) {};
\coordinate (g) at (0.5,0.7) {};
\node[arrow box] (h) at (-0.5,0.5) {$G$};
%
\draw (0,-1.4) to (p);
\draw (p) to (copier);
\draw (copier) to [out=15,in=-90] (g);
\draw (g) to [out=90,in=-90] (Ls);
\draw (h) to [out=90,in=-90] (Rs);
\draw (Ls) to (s2);
\draw (Rs) to (s3);
\draw (copier) to[out=165,in=-90] (h);
\end{tikzpicture}}}
\right)
\quad,\quad
\vcenter{\hbox{%
\begin{tikzpicture}[font=\small]
\node[state] (p) at (0,0) {$\omega$};
\node (X) at (0,0.7) {};
\draw (p) to (X);
\end{tikzpicture}}}
\right)
\\
&\xmapsto{\star}
\frac{1}{4}
\left(
\vcenter{\hbox{
\begin{tikzpicture}[font=\small,xscale=-1]
\node[state] (q) at (0,-0.1) {$\omega$};
\node[copier] (c) at (0,0.3) {};
\node[copier] (c2) at (-0.5,1.55) {};
\node[arrow box] (g) at (-0.5,0.95) {$F$};
\node[arrow box] (f) at (-1,2.25) {$G$};
\coordinate (X) at (0.7,2.85);
\coordinate (Y1) at (0,2.85);
\coordinate (Y2) at (-1,2.85);
\draw (q) to (c);
\draw (c) to[out=15,in=-90] (X);
\draw (c) to[out=165,in=-90] (g);
\draw (c2) to[out=165,in=-90] (f);
\draw (f) to (Y2);
\draw (c2) to[out=15,in=-90] (Y1);
\draw (g) to (c2);
\end{tikzpicture}}}
+
\vcenter{\hbox{
\begin{tikzpicture}[font=\small,xscale=-1]
\node[state] (q) at (0,0) {$\omega$};
\node[copier] (c) at (0,0.3) {};
\node[copier] (c2) at (-0.5,1.55) {};
\node[arrow box] (g) at (-0.5,0.95) {$F$};
\node[arrow box] (f) at (0,2.25) {$G$};
\coordinate (X) at (0.7,2.7);
\coordinate (Y1) at (0,2.7);
\coordinate (Y2) at (-1,2.7);
\coordinate (C) at (0.7,3.9);
\coordinate (B) at (0,3.9);
\coordinate (A) at (-1,3.9);
\draw (q) to (c);
\draw (c) to[out=15,in=-90] (X);
\draw (c) to[out=165,in=-90] (g);
\draw (c2) to[out=15,in=-90] (f);
\draw (f) to (Y1);
\draw (c2) to[out=165,in=-90] (Y2);
\draw (g) to (c2);
\draw (X) to [out=90,in=-90] (C);
\draw (Y1) to [out=90,in=-90] (A);
\draw (Y2) to [out=90,in=-90] (B);
\end{tikzpicture}}}
+
\vcenter{\hbox{
\begin{tikzpicture}[font=\small]
\node[state] (q) at (0,0) {$\omega$};
\node[copier] (c) at (0,0.3) {};
\node[copier] (c2) at (-0.5,1.55) {};
\node[arrow box] (g) at (-0.5,0.95) {$F$};
\node[arrow box] (f) at (0,2.25) {$G$};
\coordinate (X) at (0.7,2.95);
\coordinate (Y1) at (0,2.95);
\coordinate (Y2) at (-1,2.95);
\coordinate (C) at (0.9,4.2);
\coordinate (B) at (0,4.2);
\coordinate (A) at (-0.8,4.2);
\draw (q) to (c);
\draw (c) to[out=15,in=-90] (X);
\draw (c) to[out=165,in=-90] (g);
\draw (c2) to[out=15,in=-90] (f);
\draw (f) to (Y1);
\draw (c2) to[out=165,in=-90] (Y2);
\draw (g) to (c2);
\draw (X) to [out=90,in=-90] (A);
\draw (Y1) to [out=90,in=-90] (C);
\draw (Y2) to [out=90,in=-90] (B);
\end{tikzpicture}}}
+
\vcenter{\hbox{
\begin{tikzpicture}[font=\small]
\node[state] (p) at (0,-0.50) {$\omega$};
\node[copier] (c) at (0,-0.15) {};
\coordinate (R) at (0.7,2.6) {};
\node[arrow box] (L) at (-0.5,0.45) {$F$};
\coordinate (Ls) at (-0.8,4.0) {};
\node[copier] (c2) at (-0.5,1.05) {};
\coordinate (sL) at (-0.8,4.0) {};
\node[arrow box] (sTL) at (-1,1.8) {$G$};%
\coordinate (sTR) at (0,2.05) {};%
\coordinate (oldf) at (-1,3.0) {};
\coordinate (oldg) at (0,3.0) {};
\coordinate (TLs) at (0,4.0) {};
\coordinate (TRs) at (0.9,4.0) {};
\draw (p) to (c);
\draw (c) to [out=15,in=-90] (R);
\draw (c) to[out=165,in=-90] (L);
\draw (R) to [out=90,in=-90] (Ls);
\draw (Ls) to (sL);
\draw (L) to [out=90,in=-90] (c2);
\draw (c2) to[out=15,in=-90] (sTR);
\draw (c2) to[out=165,in=-90] (sTL);
\draw (sTL) to [out=90,in=-90] (oldg);
\draw (sTR) to [out=90,in=-90] (oldf);
\draw (oldg) to [out=90,in=-90] (TRs);
\draw (oldf) to [out=90,in=-90] (TLs);
\end{tikzpicture}}}
\right)
\end{align*}
\endgroup
By comparing these two results and using string-diagrammatic manipulations, we see that they are equal. \qedhere
\end{enumerate}
\eprf

\section{Extension to channels}
\label{sec:channelsovertime}

The graphical proof of Theorem~\ref{thm:statesovertime} is independent of whether $\omega$ is a state or a channel. 
More precisely, if one defines 
\begin{equation}
\label{eq:starmap}
\begin{split}
\hom(\mC,\mB)\times\hom(\mB,\mA)&\xrightarrow{\star}\hom(\mB\otimes\mC,\mA)\\
(G,F)&\xmapsto{\;\;\;}
G\star F:=
\frac{1}{2}\left(F\circ\bloom_{G}+F\circ\Lbloom{G}\right)\\
\left(
\vcenter{\hbox{%
\begin{tikzpicture}[font=\small]
\node[arrow box] (c) at (0,0) {$G$};
\coordinate (d) at (0,0.7) {};
\draw  (c) to (d);
\draw (c) to (0,-0.7);
\end{tikzpicture}}}
\;,\;
\vcenter{\hbox{%
\begin{tikzpicture}[font=\small]
\node[arrow box] (c) at (0,0) {$F$};
\coordinate (d) at (0,0.7) {};
\draw  (c) to (d);
\draw (c) to (0,-0.7);
\end{tikzpicture}}}
\right)
&\mapsto
\frac{1}{2}
\left(
\vcenter{\hbox{
\begin{tikzpicture}[font=\small,xscale=-1]
\node[arrow box] (p) at (0,-0.3) {$F$};
\node[copier] (copier) at (0,0.3) {};
\node[arrow box] (g) at (-0.5,0.95) {$G$};
\coordinate (X) at (0.5,1.5);
\coordinate (Y) at (-0.5,1.5);
\draw (0,-0.9) to (p);
\draw (p) to (copier);
\draw (copier) to[out=15,in=-90] (X);
\draw (copier) to[out=165,in=-90] (g);
\draw (g) to (Y);
\end{tikzpicture}}}
+
\vcenter{\hbox{
\begin{tikzpicture}[font=\small]
\node[arrow box] (p) at (0,-0.8) {$F$};
\node[copier] (copier) at (0,-0.15) {};
\coordinate (R) at (0.5,0.3) {};
\coordinate (Ls) at (-0.5,1.6) {};
\coordinate (Rs) at (0.5,1.6) {};
\coordinate (s2) at (-0.5,1.6) {};
\coordinate (s3) at (0.5,1.6) {};
\coordinate (g) at (0.5,0.7) {};
\node[arrow box] (h) at (-0.5,0.5) {$G$};
%
\draw (0,-1.4) to (p);
\draw (p) to (copier);
\draw (copier) to [out=15,in=-90] (g);
\draw (g) to [out=90,in=-90] (Ls);
\draw (h) to [out=90,in=-90] (Rs);
\draw (Ls) to (s2);
\draw (Rs) to (s3);
\draw (copier) to[out=165,in=-90] (h);
\end{tikzpicture}}}
\right)
\end{split}
\end{equation}
(bypassing the Choi--Jamio{\l}kowski isomorphism altogether), 
then the proof of Theorem~\ref{thm:statesovertime} goes through without any changes. 
This seems to go against the no-go theorems of~\cite{HHPBS17}. The resolution to this seeming paradox comes from the `preservation of classical limit' axiom. We have shown that our formulation of this axiom is equivalent to the one of~\cite{HHPBS17} when dealing with turning a channel plus state into a joint state (cf.\ Proposition~\ref{prop:quantumclassical}). However, when extending our definition of a classical model (Definition~\ref{defn:classicalmodel}) to channels, as opposed to just states, we find that our definition is inequivalent to the axiom of commutativity enforced in the no-go theorems of~\cite{HHPBS17}. Thus, by using categorical reasoning and the framework of quantum Markov categories~\cite{PaBayes}, we are able to bypass the no-go theorems of~\cite{HHPBS17}. The present section will illustrate how this works. 

\begin{notation}
In all definitions made before, wherever a linear functional $\omega:\mA\to\C$ appears, the same exact definition is now made for a linear map $\omega:\mA\to\mZ$, where $\mZ$ is some finite-dimensional $C^*$-algebra. For example, a \define{classical model} for a pair $(\mB\xrightarrow{F}\mA,\mA\xrightarrow{\omega}\mZ)$ consists of commutative $C^*$-algebras $j_{\mA}:\mA_{\mathrm{cl}}\hookrightarrow\mA$, $j_{\mB}:\mB_{\mathrm{cl}}\hookrightarrow\mB$, conditional expectations $E_{\mA}:\mA\to\mA_{\mathrm{cl}}$, $E_{\mB}:\mB\to\mB_{\mathrm{cl}}$, and a linear map $F_{\mathrm{cl}}:\mB_{\mathrm{cl}}\to\mA_{\mathrm{cl}}$ such that%
\footnote{As in the case of states, $\omega\circ F=(\omega\circ F)_{\restriction}\circ E_{\mB}$ is a consequence of the other two conditions.}
\[
F=j_{\mA}\circ F_{\mathrm{cl}}\circ E_{\mB}, \qquad
\omega=\omega_{\restriction}\circ E_{\mA}, 
\qquad
\text{ and }
\qquad
\omega\circ F=(\omega\circ F)_{\restriction}\circ E_{\mB}, 
\]
where the $\restriction$ subscript means restriction to the commutative subalgebras. 
The only difference in notation/terminology between this section and the previous sections is that a `family of states over time function' is replaced with a `family of channels over time function.' Again, this is slightly abusive terminology since channels here need not be positive. Note, however, that the compositionality/associativity axiom can now be formulated much more simply without ever even using the Choi--Jamio{\l}kowski isomorphism. Namely, it says that the diagram 
\[
\xy0;/r.22pc/:
(-43,10)*+{\hom(\mC,\mB)\times\hom(\mB,\mA)\times\hom(\mA,\mZ)}="1";
(43,10)*+{\hom(\mB\otimes\mC,\mA)\times\hom(\mA,\mZ)}="2";
(-43,-10)*+{\hom(\mC,\mA\otimes\mB)\times\hom(\mA\otimes\mB,\mZ)}="3";
(43,-10)*+{\hom(\mA\otimes\mB\otimes\mC,\mZ)}="4";
{\ar"1";"2"^(0.55){\star\times\id}};
{\ar"1";"3"_{(\shriek_{\mA}\otimes\;\cdot\;)\times\star}};
{\ar"2";"4"^{\star}};
{\ar"3";"4"_(0.55){\star}};
\endxy
\]
commutes, i.e., 
\[
(G\star F)\star\omega=(\shriek_{A}\otimes G)\star(F\star\omega).
\]
\end{notation}

\bn
\label{prop:classicalmodelequiv}
Given $\dag$-preserving linear maps $(\mB\xrightarrow{F}\mA,\mA\xrightarrow{\omega}\mZ)$, the following are equivalent.
\begin{enumerate}[i.]
\itemsep0em
\item
\label{item:wblcbl}
The identity $\omega\circ\bloom_{F}=\omega\circ\Lbloom{F}$ holds.
\item
\label{item:dencom}
The densities $1_{\mZ}\otimes\Cd[F]$ and $\Cd[\omega]\otimes1_{\mB}$ commute, i.e., 
$\big[1_{\mZ}\otimes\Cd[F],\Cd[\omega]\otimes1_{\mB}\big]=0$.
\end{enumerate}
Furthermore, if the pair $(F,\omega)$ admits a classical model, then items~\ref{item:wblcbl} and~\ref{item:dencom} hold.
\en

\br
Note that when $\mZ=\C$, Proposition~\ref{prop:classicalmodelequiv} provides a strengthening of Proposition~\ref{prop:quantumclassical}. In addition, Proposition~\ref{prop:classicalmodelequiv} illustrates in what sense a certain commutativity condition holds when a system is effectively classical. This commutativity condition is what replaces the commutativity condition in~\cite{HHPBS17} and allows us to guarantee that a family of channels over time function exists (see Theorem~\ref{thm:channelsovertime} below).
\er

\blem
\label{lem:HSmult}
Let $\mA=\bigoplus_{x\in X}\matr_{m_{x}}(\C)$. Then 
\[
\bigoplus_{x\in X}\matr_{m_{x}}(\C)\xrightarrow{\mu_{\mA}^{*}}\bigoplus_{x',x''\in X}\left(\matr_{m_{x'}}(\C)\otimes\matr_{m_{x''}}(\C)\right),
\]
the Hilbert--Schmidt adjoint of the multiplication map, is given explicitly by the formula 
\[
(\mu_{\mA}^*)_{(x',x'')x}(A)=\delta_{xx'}\delta_{xx''}\sum_{i,j,k}A_{ij}E_{ik}^{(m_{x})}\otimes E_{kj}^{(m_{x})}
\]
for all $A\in\matr_{m_{x}}(\C)$ and for all $x,x',x''\in X$. Here, $A_{ij}$ denotes the $ij$-th entry of $A$ with respect to the standard basis. 
\elem

\bprf
[Proof of Lemma~\ref{lem:HSmult}]
This follows from the fact that $(\mu_{\mA})_{x(x',x'')}$ vanishes unless $x=x'=x''$, the definition of the Hilbert--Schmidt inner product, and 
the fact that multiplication is computed component-wise. 
\eprf

The following lemma generalizes Lemma~\ref{lem:jordanbloomdensities}. 

\blem
\label{lem:cdbloom}
Given linear maps $(\mB\xrightarrow{F}\mA,\mA\xrightarrow{\omega}\mZ)$, 
\[
\Cd[\omega\circ\bloom_{F}]
=\big(1_{\mZ}\otimes\Cd[F]\big)\big(\Cd[\omega]\otimes1_{\mB}\big)
\quad\text{ and }\quad
\Cd[\omega\circ\Lbloom{F}]
=\big(\Cd[\omega]\otimes1_{\mB}\big)\big(1_{\mZ}\otimes\Cd[F]\big).
\]
\elem

\bprf
[Proof of Lemma~\ref{lem:cdbloom}]
By the distributive property of $\otimes$ and $\oplus$ along with Lemma~\ref{lem:HSmult}, it suffices to assume the algebras are matrix algebras. 
Set $\omega^{ij}_{\beta\alpha}\in\C$ to be the unique numbers satisfying $\omega^*(E_{\beta\alpha})=\sum_{i,j}\omega^{ij}_{\beta\alpha}E_{ij}$ for all $\alpha,\beta$. Then 
\[
\begin{split}
\Cd[\omega\circ\bloom_{F}]^{\dag}
&=\Big(\id_{\mZ}\otimes\big(\omega\circ\mu_{\mA}\circ(\id_{\mA}\otimes F)\big)^{*}\Big)\left(\sum_{\alpha,\beta}E_{\alpha\beta}\otimes E_{\beta\alpha}\right)\\
&=(\id_{\mZ}\otimes\id_{\mA}\otimes F^{*})\left(\sum_{\alpha,\beta}E_{\alpha\beta}\otimes\mu_{\mA}^{*}\big(\omega^*(E_{\beta\alpha})\big)\right)\\
&=(\id_{\mZ}\otimes\id_{\mA}\otimes F^{*})\left(\sum_{\alpha,\beta}\sum_{i,j}\omega^{ij}_{\beta\alpha}E_{\alpha\beta}\otimes\mu_{\mA}^{*}(E_{ij})\right)\\
&=\sum_{\alpha,\beta}\sum_{i,j,k}\omega^{ij}_{\beta\alpha} E_{\alpha\beta}\otimes E_{ik}\otimes F^{*}(E_{kj}).
\end{split}
\]
Meanwhile, 
\[
\begin{split}
\Big(\big(1_{\mZ}\otimes\Cd[F]\big)\big(\Cd[\omega]\otimes1_{\mB}\big)\Big)^{\dag}
&=\sum_{\alpha,\beta}\sum_{i,j}E_{\alpha\beta}\otimes\omega^*(E_{\beta\alpha})E_{ij}\otimes F^{*}(E_{ji})\\
&=\sum_{\alpha,\beta}\sum_{i,j,k,l}\omega^{kl}_{\beta\alpha}E_{\alpha\beta}\otimes(\underbrace{E_{kl}E_{ij}}_{\delta_{li}E_{kj}})\otimes F^{*}(E_{ji})\\
&=\sum_{\alpha,\beta}\sum_{i,j,k}\omega^{ki}_{\beta\alpha} E_{\alpha\beta}\otimes E_{kj}\otimes F^{*}(E_{ji}).
\end{split}
\]
By relabelling the dummy indices, the two expressions are seen to be the same. Applying $\dag$ to both sides proves the first identity. 
The other identity follows from similar calculations. 
\eprf

\bprf
[Proof of Proposition~\ref{prop:classicalmodelequiv}]
The proof of the last claim implies item~\ref{item:wblcbl} follows exactly the same argument as in the proof of the `preservation of classical limit' in Theorem~\ref{thm:statesovertime}. The equivalence between item~\ref{item:wblcbl} and item~\ref{item:dencom} follows from Lemma~\ref{lem:cdbloom}. 
\eprf

\bt
\label{thm:channelsovertime}
A family of channels over time function exists and (\ref{eq:starmap}) provides an explicit construction.
\et

\bprf
The proof is completely analogous to the proof of Theorem~\ref{thm:statesovertime}.
\eprf

\section{Discussion}
\label{sec:discussion}

In this paper, we constructed a consistent way of associating a joint `state' on $\mA\otimes\mB$ with every state on $\mA$ and a quantum channel%
\footnote{In the Heisenberg picture, the directionality of the arrows is $\mB\to\mA$ and is the convention followed in the present paper.} 
$\mA\to\mB$, in such a way that by-passes the no-go result of~\cite{HHPBS17}. The reason `state' is in quotes is because the associated joint matrix is only self-adjoint in general, but is not necessarily positive. Therefore, it remains an open question whether there exists a consistent manner of associating a genuinely positive joint state to an initial (positive) state and a positive (perhaps even completely positive) map. 
In particular, we do not know whether there is such an assignment satisfying the axioms we have outlined that also includes such a positivity constraint.  
Furthermore, although we have provided a construction of a family of states over time function, we have made no claim as to the uniqueness of such an assignment. In particular, we do not know if the Jordan product provides the unique function that satisfies these axioms.

An interesting aspect of our work is that the proof of the main theorem was provided in the setting of (enriched) quantum Markov categories~\cite{PaBayes}. The proof itself also illustrated a natural generalization to channels, where the `preservation of the classical limit' axiom of~\cite{HHPBS17} was replaced by an alternative one that allowed us to bypass the no-go result of~\cite{HHPBS17}. It seems reasonable to suspect that extensions to certain von~Neumann algebras are possible, though this is only a speculation. We leave this question to the interested reader. 

Yet another question that arises as a result of our theorem is related to quantum conditionals, which can be viewed as the opposite procedure to the one described in this work. In particular, if one is given a joint state, can one find a process for which the joint state can be expressed in terms of this process and its marginal? In~\cite{PaQPL21}, it was shown that one can express a joint state $\zeta$ on $\mA\otimes\mB$ as $\zeta=\omega\circ\bloom_{F}$ for some \emph{positive} $F$, and where $\omega=\zeta\circ i_{\mA}$, if and only if some non-trivial condition holds. The results of this paper suggest that perhaps one should change the question to the existence of a positive $F$ such that $\zeta=\frac{1}{2}(\omega\circ\bloom_{F}+\omega\circ\Lbloom{F})$. It is not presently known if such a symmetrization procedure allows more conditionals to exist. 

Finally, it would be interesting if our state over time function can be used to define information measures associated with a quantum channel and an initial state. In particular, if a joint entropy may be associated with our state over time, it would be straightforward to define an associated conditional entropy and mutual information by mimicking the defining formulas in the classical case. Moreover, if such a joint entropy exists, it would be worth investigating whether or not such a joint entropy yields quantum analogues of the characterizations of classical information measures appearing in~\cite{FuPa21}.

\bigskip
\noindent
{\bf Acknowledgements.}
The authors thank Robert W.\ Spekkens for answering their questions regarding~\cite{HHPBS17}, Chris Heunen for helpful suggestions, and Tobias Fritz for informing us about the references~\cites{Ga96,CoGa99}. The authors also thank the anonymous referees who provided several key observations and suggestions. This work is supported by MEXT-JSPS Grant-in-Aid for Transformative Research Areas (A) ``Extreme Universe'', No.\ 21H05183. A majority of this work was completed while AJP was at the Institut des Hautes \'Etudes Scientifiques.

\addcontentsline{toc}{section}{\numberline{}Bibliography}
\bibliographystyle{plain}
\bibliography{references}

\begin{bibdiv}
\begin{biblist}

\bib{Bo51}{book}{
      author={Bohm, David},
       title={{Quantum theory}},
   publisher={Prentice-Hall},
     address={Englewood Cliffs, {NJ}},
        date={1951},
        note={Also as reprint ed.: New York, NY, Dover Publications, 1989},
}

\bib{ChJa18}{article}{
      author={Cho, Kenta},
      author={Jacobs, Bart},
       title={Disintegration and {B}ayesian inversion via string diagrams},
        date={2019},
     journal={Math. Struct. Comp. Sci.},
       pages={1\ndash 34},
      eprint={1709.00322},
         url={https://doi.org/10.1017/S0960129518000488},
}

\bib{Ch75}{article}{
      author={Choi, Man~Duen},
       title={Completely positive linear maps on complex matrices},
        date={1975},
     journal={Linear Algebra Its Appl.},
      volume={10},
       pages={285\ndash 290},
         url={https://doi.org/10.1016/0024-3795(75)90075-0},
}

\bib{CDDG17}{inproceedings}{
      author={Clerc, Florence},
      author={Danos, Vincent},
      author={Dahlqvist, Fredrik},
      author={Garnier, Ilias},
       title={Pointless learning},
organization={Springer},
        date={2017},
   booktitle={International conference on foundations of software science and
  computation structures},
       pages={355\ndash 369},
         url={https://doi.org/10.1007/978-3-662-54458-7_21},
}

\bib{CoKi17}{book}{
      author={Coecke, Bob},
      author={Kissinger, Aleks},
       title={Picturing quantum processes: A first course in quantum theory and
  diagrammatic reasoning},
   publisher={Cambridge University Press},
        date={2017},
}

\bib{CoGa99}{article}{
      author={Corradini, Andrea},
      author={Gadducci, Fabio},
       title={An algebraic presentation of term graphs, via gs-monoidal
  categories},
        date={1999},
     journal={Appl. Categ. Structures},
      volume={7},
      number={4},
       pages={299\ndash 331},
}

\bib{EPR}{article}{
      author={Einstein, Albert},
      author={Podolsky, Boris},
      author={Rosen, Nathan},
       title={Can quantum-mechanical description of physical reality be
  considered complete?},
        date={1935},
     journal={Phys. Rev.},
      volume={47},
       pages={777\ndash 780},
         url={https://link.aps.org/doi/10.1103/PhysRev.47.777},
}

\bib{FJV15}{article}{
      author={Fitzsimons, Joseph~F.},
      author={Jones, Jonathan~A.},
      author={Vedral, Vlatko},
       title={Quantum correlations which imply causation},
        date={2015},
     journal={Sci. Rep.},
      volume={5},
      number={1},
       pages={18281},
      eprint={1302.2731},
         url={https://doi.org/10.1038/srep18281},
}

\bib{Fr20}{article}{
      author={Fritz, Tobias},
       title={A synthetic approach to {M}arkov kernels, conditional
  independence and theorems on sufficient statistics},
        date={2020},
     journal={Adv. Math.},
      volume={370},
       pages={107239},
      eprint={1908.07021},
         url={https://doi.org/10.1016/j.aim.2020.107239},
}

\bib{FuPa21}{article}{
      author={Fullwood, James},
      author={Parzygnat, Arthur~J.},
       title={The information loss of a stochastic map},
        date={2021},
        ISSN={1099-4300},
     journal={Entropy},
      volume={23},
      number={8},
      eprint={2107.01975},
         url={https://www.mdpi.com/1099-4300/23/8/1021},
}

\bib{Ga96}{thesis}{
      author={Gadducci, Fabio},
       title={On the algebraic approach to concurrent term rewriting},
        type={Ph.D. Thesis},
        date={1996},
  note={\url{http://citeseerx.ist.psu.edu/viewdoc/summary?doi=10.1.1.57.5651}},
}

\bib{GPRR21}{misc}{
      author={{Giorgetti}, Luca},
      author={{Parzygnat}, Arthur~J.},
      author={{Ranallo}, Alessio},
      author={{Russo}, Benjamin~P.},
       title={{Bayesian inversion and the Tomita-Takesaki modular group}},
        date={2021},
        note={arXiv preprint:
  \href{https://arxiv.org/abs/2112.03129}{2112.03129 [math.OA]}},
}

\bib{GPS21}{article}{
      author={{Girard}, Mark},
      author={{Pl{\'a}vala}, Martin},
      author={{Sikora}, Jamie},
       title={{Jordan products of quantum channels and their compatibility}},
        date={2021},
     journal={Nature Communications},
      volume={12},
       pages={2129},
      eprint={2009.03279},
}

\bib{HeVi19}{book}{
      author={Heunen, Chris},
      author={Vicary, Jamie},
       title={Categories for quantum theory: an introduction},
   publisher={Oxford University Press, USA},
        date={2019},
}

\bib{HHPBS17}{article}{
      author={{Horsman}, Dominic},
      author={{Heunen}, Chris},
      author={{Pusey}, Matthew~F.},
      author={{Barrett}, Jonathan},
      author={{Spekkens}, Robert~W.},
       title={{Can a quantum state over time resemble a quantum state at a
  single time?}},
        date={2017},
     journal={Proc. R. Soc. A},
      volume={473},
      number={2205},
       pages={20170395},
      eprint={1607.03637},
}

\bib{Ja72}{article}{
      author={Jamio{\l}kowski, Andrzej},
       title={Linear transformations which preserve trace and positive
  semidefiniteness of operators},
        date={1972},
        ISSN={0034-4877},
     journal={Rep. Math. Phys.},
      volume={3},
      number={4},
       pages={275\ndash 278},
         url={https://doi.org/10.1016/0034-4877(72)90011-0},
}

\bib{Kr83}{book}{
      author={Kraus, Karl},
       title={States, effects, and operations},
    subtitle={Fundamental notions of quantum theory},
      series={Lecture Notes in Physics},
   publisher={Springer Berlin Heidelberg},
        date={1983},
         url={https://doi.org/10.1007/3-540-12732-1},
}

\bib{Le06}{article}{
      author={Leifer, Matthew~S.},
       title={Quantum dynamics as an analog of conditional probability},
        date={2006},
     journal={Phys. Rev. A},
      volume={74},
       pages={042310},
      eprint={0606022},
         url={https://link.aps.org/doi/10.1103/PhysRevA.74.042310},
}

\bib{Le07}{inproceedings}{
      author={{Leifer}, Matthew~S.},
       title={{Conditional Density Operators and the Subjectivity of Quantum
  Operations}},
        date={2007},
   booktitle={Foundations of probability and physics - 4},
      editor={{Adenier}, Guillaume},
      editor={{Fuchs}, Chrisopher},
      editor={{Khrennikov}, Andrei~Yu},
      series={American Institute of Physics Conference Series},
      volume={889},
       pages={172\ndash 186},
}

\bib{LeSp13}{article}{
      author={Leifer, Matthew~S.},
      author={Spekkens, Robert~W.},
       title={Towards a formulation of quantum theory as a causally neutral
  theory of {B}ayesian inference},
        date={2013},
     journal={Phys. Rev. A},
      volume={88},
       pages={052130},
      eprint={1107.5849},
         url={https://link.aps.org/doi/10.1103/PhysRevA.88.052130},
}

\bib{PaBayes}{misc}{
      author={Parzygnat, Arthur~J.},
       title={Inverses, disintegrations, and {B}ayesian inversion in quantum
  {M}arkov categories},
        date={2020},
        note={arXiv preprint:
  \href{https://arxiv.org/abs/2001.08375}{2001.08375 [quant-ph]}},
}

\bib{PaQPL21}{inproceedings}{
      author={Parzygnat, Arthur~J.},
       title={Conditional distributions for quantum systems},
        date={2021},
   booktitle={Proceedings 18th {I}nternational {C}onference on {Q}uantum
  {P}hysics and {L}ogic},
      series={Electronic Proceedings in Theoretical Computer Science (EPTCS)},
      volume={343},
       pages={1\ndash 13},
}

\bib{ZPTGVF18}{article}{
      author={Zhao, Zhikuan},
      author={Pisarczyk, Robert},
      author={Thompson, Jayne},
      author={Gu, Mile},
      author={Vedral, Vlatko},
      author={Fitzsimons, Joseph~F.},
       title={Geometry of quantum correlations in space-time},
        date={2018},
     journal={Phys. Rev. A},
      volume={98},
       pages={052312},
      eprint={1711.05955},
         url={https://link.aps.org/doi/10.1103/PhysRevA.98.052312},
}

\end{biblist}
\end{bibdiv}

\Addresses

\end{document}